\documentclass[aps,prd,twocolumn,showpacs,groupedaddress,amsmath,amssymb, superscriptaddress,nofootinbib]{revtex4-2} 
\usepackage[dvipsnames]{xcolor}

\usepackage{graphicx}
\usepackage{dcolumn}
\usepackage{multirow} 
\usepackage{bm}
\usepackage[colorlinks,linkcolor=blue,urlcolor=blue,citecolor=blue]{hyperref}

\usepackage{color}
\usepackage{soul}
\usepackage[utf8]{inputenc}
\usepackage[bottom]{footmisc}
\usepackage[T1]{fontenc}
\usepackage{mathptmx}
\usepackage{subcaption}
\usepackage{url}

\newcommand\unit[1]{{\rm #1}}

\newcommand\prr{Phys. Rev. Res.}
\newcommand\apjl{Astrophys. J. Lett.}
\newcommand\apjs{Astrophys. J. Suppl. Ser.}
\newcommand\cqg{Class. Quantum Grav.}
\newcommand\mnras{Mon. Not. R. Astron. Soc.}
\newcommand\aap{Astron. Astrophys.}
\newcommand\pasj{Publ. Astron. Soc. Jpn.}
\newcommand\lrr{Living Rev. Relativ.}
\newcommand\grg{Gen. Relativ. Gravit.}
\newcommand\nata{Nat. Astron.}

\def\RIT{Center for Computational Relativity and Gravitation, Rochester Institute of Technology, Rochester, New York 14623, USA}
\def\XCP{Computational Physics Division, Los Alamos National Laboratory, Los Alamos, New Mexico 87545, USA}
\def\CTA{Center for Theoretical Astrophysics, Los Alamos National Laboratory, Los Alamos, New Mexico 87545, USA}
\def\CCSS{Computer, Computational, and Statistical Sciences Division, Los Alamos National Laboratory, Los Alamos, New Mexico 87545, USA}
\def\TD{Theoretical Division, Los Alamos National Laboratory, Los Alamos, New Mexico 87545, USA}
\def\ISR{Intelligence and Space Research Division, Los Alamos National Laboratory, Los Alamos, New Mexico 87545, USA}
\def\UA{University of Arizona, Tucson, Arizona 85721, USA}
\def\NM{Department of Physics and Astronomy, University of New Mexico, Albuquerque, New Mexico 87131, USA}
\def\GW{George Washington University, Washington, DC 20052, USA}

\begin{document}

\title{Surrogate light curve models for kilonovae with comprehensive wind ejecta outflows \\ and parameter estimation for AT2017gfo}

\author{Atul Kedia}
\author{Marko Ristic}
\author{Richard O'Shaughnessy}
\author{Anjali B. Yelikar}
\affiliation{\RIT}
\author{Ryan T. Wollaeger}
\affiliation{\CTA}
\affiliation{\CCSS}
\author{Oleg Korobkin}
\affiliation{\CTA}
\affiliation{\TD}
\author{Eve A. Chase}
\affiliation{\CTA}
\affiliation{\ISR}
\author{Christopher L. Fryer}
\affiliation{\CTA}
\affiliation{\CCSS}
\affiliation{\UA}
\affiliation{\NM}
\affiliation{\GW}
\author{Christopher J. Fontes}
\affiliation{\CTA}
\affiliation{\XCP}

\date{\today}

\begin{abstract}

The electromagnetic emission resulting from neutron star mergers have been shown to encode properties of the ejected material in their light curves. The ejecta properties inferred from the kilonova emission has been in tension with those calculated based on the gravitational wave signal and numerical relativity models. Motivated by this tension, we construct a broad set of surrogate light curve models derived for kilonova ejecta. The four-parameter family of two-dimensional anisotropic simulations and its associated surrogate explore different assumptions about the wind outflow morphology and outflow composition, keeping the dynamical ejecta component consistent. We present the capabilities of these surrogate models in interpolating kilonova light curves across various ejecta parameters and perform parameter estimation for AT2017gfo both without any assumptions on the outflow and under the assumption that the outflow must be representative of solar \emph{r}-process abundance patterns. Our parameter estimation for AT2017gfo shows these surrogate models help alleviate the ejecta property discrepancy while also illustrating the impact of systematic modeling uncertainties on these properties, urging further investigation.
  
\end{abstract}

\maketitle

\section{Introduction}

Merging neutron stars have been demonstrated to emit both gravitational waves (GW) and a variety of accessible electromagnetic counterparts, as shown by the observation of GW170817 and the following transient AT2017gfo \cite{2017ApJ...848L..12A, LIGO-GW170817-bns, Tanvir_2017, 2017Natur.551...75S, 2017ApJ...848L..17C, 2017Natur.551...67P, 2017Natur.551...64A}. Also emitted during merger is the most unambiguous indication of matter in these systems:  nuclear matter ejected due to the merger itself, which over time expands and heats through ongoing radioactive decay, producing a distinctive “kilonova” emission \cite{1974ApJ...192L.145L, 1998ApJ...507L..59L, metzger2020kilonovae, 2020GReGr..52..108B}. 
Particularly in conjunction with gravitational wave observations \cite{2017ApJ...848L..12A, LIGO-GW170817-bns, LIGO-GW170817-H0, LIGO-GW170817-EOS, LIGO-GW170817-EOSrank, 2020PhRvL.125n1103B, 2019NatAs...3..940H, 2020Sci...370.1450D}, 
kilonova discoveries can provide insight into the physics and significance of this radioactive ejecta. On the one hand, these counterparts probe uncertain nuclear physics \cite{2021ApJ...918...44B,2021ApJ...906...94Z, 2020arXiv200604322V, 2019AnPhy.41167992H, 2020GReGr..52..109C}. On the other, these processes may be in part responsible for the production of \emph{r}-process elements throughout the universe. \cite{1974ApJ...192L.145L, 2017Sci...358.1570D, 2021ApJ...918...44B}.

Kilonova observations in principle encode the amount and properties of the ejected material in multiwavelength light curves (LCs) and spectra \cite{metzger2020kilonovae,2018MNRAS.480.3871C,2017ApJ...851L..21V}.
Previous investigations have characterized this ejected material via two components, denoted as dynamical and disk wind ejecta, reflecting differences in their formation and ejection mechanisms. 
Several studies of GW170817 attempted to infer the amount of material ejected \cite{2018MNRAS.480.3871C,2017ApJ...851L..21V,Breschi2021AT2017gfo,gwastro-mergers-em-CoughlinGPKilonova-2020, 2019MNRAS.489L..91C, 2017Natur.551...75S, Tanvir_2017, 2017ApJ...848L..21A, chornock17, 2017ApJ...848L..17C,2020ApJ...889..171K,Ristic22, collins20223d}. However, the amount and properties of the ejected material from this event remain uncertain and in considerable tension with theoretical expectations for the amount of each type of ejecta \cite{2021ApJ...906...98N,2020ARNPS..7013120R, 2021ApJ...910..116K, Ristic22}, likely in part because of underestimated uncertainties in these theoretical ejecta estimates \cite{henkel2022study}.

The tension between observation and expectations could in principle reflect modeling systematic errors. These observations have historically been interpreted with semianalytic models \cite{1998ApJ...507L..59L,metzger2020kilonovae,2014MNRAS.439..757G} as they can be evaluated quickly and continuously over the parameters, which characterize potential merger ejecta. However, it is well known that these semianalytic models contain oversimplified physics of already simplified radiative transfer calculations \cite{2018MNRAS.478.3298W,2020ApJ...899...24E, kilonova-lanl-WollaegerNewGrid2020} that neglect detailed full three-dimensional anisotropic radiative transfer, opacity, sophisticated nuclear reaction networks, and composition differences.
That said, recent calculations using improved anisotropic radiative transfer and opacity calculations still arrive at qualitatively similar conclusions for AT2017gfo \cite{Ristic22, gwastro-mergers-em-CoughlinGPKilonova-2020,kawaguchi2018radiative,2020ApJ...889..171K,2021arXiv211215470A, collins20223d}: a relatively high mass blue component outflowing along the poles, and a significant mass in a red component outflowing preferentially towards the equator. Reference \cite{Ristic22} also identifies a larger velocity for the blue component, although the result reverses upon excluding bands that exhibit large uncertainties. However, recent numerical relativity simulations give surprisingly low dynamical outflow velocities when mass averaged, e.g., see Ref. \cite{2021ApJ...906...98N} for example, and are not in as much disagreement with regards to the red component. (Some recent calculations with more complex outflow morphologies, however, recover wind speeds more consistent with prior expectations \cite{Breschi2021AT2017gfo}.)
Essentially, the best-fitting properties of the ejected material in our polar (blue) and equatorial (red) ejecta are in tension with theoretical expectations for their physical origin.

Motivated by this tension, in this paper we revisit our detailed anisotropic radiative transfer calculations and inference \cite{kilonova-lanl-WollaegerNewGrid2020, Ristic22}, now allowing for a wide selection of outflow morphologies \cite{2021ApJ...910..116K} to see impact on ejecta properties of AT2017gfo. The properties of the outflow that we vary in this paper are the outflow mass-density profile for each component (referred to as the ``morphology'' for the rest of the paper) and the nuclear composition using the electron fraction $Y_e \equiv (n_p)/(n_p + n_n)$ as a proxy, where $n_p$ is the number of protons and $n_n$ is the number of neutrons in the ejecta. Effects of different dynamical ejecta morphologies have been studied in the past to show significant light curve (LC) shifts \cite{2021ApJ...910..116K}, with a larger dependence on morphologies as compared to the anecdotal ejecta masses and velocities. Certain outflow models create a strong angular differential in the LC by suppressing the bluer bands along the equatorial plane (also known as lanthanide curtaining) \cite{barnes2013effect, 2015MNRAS.450.1777K, 2021MNRAS.500.1772N}.

In this paper we present a new set of surrogate models describing kilonova LCs resulting from a broad set of neutron star merger ejecta. This paper is organized as follows.
In Sec. \ref{sec:method} we review our approach of adaptively generating kilonovae simulations, constructing the surrogate LC model using Gaussian process regression, and performing parameter estimation (PE) for ejecta properties. We also discuss the specific alternate outflow morphologies and compositions employed in this paper. For context and to better highlight current tension with contemporary theoretical models, we also introduce estimates for the ejected material in each component applied to interpreting GW170817 given the well-constrained source distance and inclination.
In Sec. \ref{sec:results} we discuss the new surrogate LC models and describe our PE results obtained using them. We further conclude this section with a discussion on the implications of this tension for joint multimessenger inference with contemporary ejecta models.
We then conclude in Sec. \ref{sec:conclusions}.
We provide all the simulation models, the surrogate models developed and employed in this paper, and sample codes to generate LCs at a public Zenodo repository\footnote{\label{note1}\url{https://zenodo.org/record/7335961}}. The surrogates and sample codes are also available at a public GitHub repository\footnote{\label{note2}\url{https://github.com/markoris/surrogate_kne}}.

\section{Method}
\label{sec:method}

The workflow pipeline is discussed in detail in this section and can be summarized as follows: We start by generating initial simulations for emitted spectra using the radiative transfer code \texttt{SuperNu}. Then our active learning scheme iteratively chooses the ``best'' ejecta parameters to simulate until the simulation space is sufficiently large in size to produce a robust surrogate model. We then construct high-fidelity multidimensional surrogate LC models for each morphology and composition combination using Gaussian process regression (GP) as was performed in \cite{Ristic22}. These surrogate models are now capable of making predictions of LCs along several model parameters. The surrogate models are then used to infer parameters for AT2017gfo and we compare these electromagnetically inferred ejecta parameters with inferred ejecta parameters from the corresponding gravitational wave signal GW170817 using a library of numerical relativity-based ejecta models. These steps are further discussed and detailed in the following subsections.

\subsection{Simulation methods and placement}
\label{sec:simulation}
We perform simulations of kilonova LCs using the time-dependent radiative transfer code \texttt{SuperNu} \cite{wollaeger2013radiation, 2014ApJS..214...28W}, following the methodology presented in \cite{2018MNRAS.478.3298W, Ristic22, kilonova-lanl-WollaegerNewGrid2020, 2021ApJ...910..116K} and references therein. To determine nuclear heating rates, we use the detailed, time-dependent radioactive isotope composition results from the nucleosynthesis simulations with the \texttt{WinNet} code \cite{2012ApJ...750L..22W, Korobkin_2012} and include contributions of individual radiation species (such as $\alpha$-, $\beta$-, $\gamma$-radiation, and fission products) for each isotope. The contributions are then weighted by thermalization efficiencies from \cite{Barnes_2016} (a detailed description of the adopted nuclear heating can be seen in Ref. \cite{2018MNRAS.478.3298W} ). These heating rates and composition effects together along with the tabulated binned opacities resulting from the the Los Alamos suite of atomic physics codes \cite{2015JPhB...48n4014F,2020MNRAS.493.4143F}, give the resultant kilonova LCs. The tabulated binned opacities used in this paper, however, are not calculated for all elements. Instead we produce opacities from representative proxy elements by combining pure-element opacities of nuclei with similar atomic properties \cite{2020MNRAS.493.4143F}. The output we obtain from \texttt{SuperNu} are light spectra for 54 viewing angle bins uniformly distributed between 0 and 180 deg in cosine space. For each viewing angle bin, the light rays are collected from across the face of the ejecta as seen from the direction of that viewing angle.

To simplify comparison with previous paper, we adopted the same set of representative elements as in past paper \cite{kilonova-lanl-WollaegerNewGrid2020, Ristic22}, however, a recent study \cite{fontes2022actinide} includes a more comprehensive list of elements, including actinides. A comparison in Fig. 8 in Ref. \cite{fontes2022actinide} shows that when the actinide opacities are improved upon the luminosity varies by $\sim 15\%$ at its maximum displacement. As will be discussed later in Sec. \ref{sec:results}, this displacement is currently significantly smaller than our GP interpolation error making the changes with the new elemental approach less significant for the purposes of this paper. However, any followup work should utilize the complete set of elemental opacities.

\subsection{Initial conditions: Wind morphologies and compositions}
As initial conditions for the \texttt{SuperNu} code, we employ a two-component ejecta model, comprising an unbound dynamical ejecta from the merger, followed by the disk wind ejecta. All ejecta components are assumed to be expanding homologously with a prescribed velocity profile, mass distribution versus angle (see Ref. \cite{2021ApJ...910..116K} for their complete description), and prescribed uniform composition set by its initial electron fraction $Y_e$. As discussed earlier, we use $Y_e$ as a proxy for ejecta composition in these simulations.

We simulate a range of wind ejecta morphologies and compositions in order to study their influence on kilonova LCs, and build robust surrogate models that can be used for parameter inference of kilonovae observations. The outflow morphologies we model are selected from previously studied models \cite{2021ApJ...910..116K, kilonova-lanl-WollaegerNewGrid2020} with a varying set of wind ejecta structure and electron fraction. Table \ref{table:models} shows the morphology and composition combinations simulated. Here, for instance, the model labeled TPwind1 comprises a Torus shaped dynamical ejecta, and a Peanut shaped wind ejecta (S would indicate a Spherical wind ejecta) with their mathematical form given in Ref. \cite{2021ApJ...910..116K}. The labels ``wind1'' and ``wind2'' denote the two choices of wind ejecta composition. wind1 corresponds to the less neutron rich configuration with $Y_e = 0.37$ whereas wind2 corresponds to a more neutron rich $Y_e = 0.27$; wind2 with a Peanut morphology, i.e., TPwind2, was adopted in the previous paper \cite{Ristic22} and hence a new surrogate model for the case is not generated in this study. As was in previous paper, we fix the composition of the dynamical ejecta as $Y_e = 0.04$.

\begin{table}[tbpt]
\caption{Ejecta morphologies and compositions studied in this paper. The composition of the dynamical component is fixed at $Y_e$ = 0.04. In terms of this notation, the previous investigation studied a TPwind2 outflow \cite{Ristic22}. }
\begin{tabular}{@{\extracolsep{25pt}} l c c c @{}}
    \hline \hline
    \multirow{2}{*}{Name} &
    \multicolumn{2}{c}{Wind} & \multirow{2}{*}{Dynamical} \\\cline{2-3} &  Morphology & $Y_e$
    \\ \hline
    TPwind1 &  Peanut & $ 0.37$ & Torus \\
    TSwind1 &  Spherical & $ 0.37$ & Torus  \\
    TSwind2 &  Spherical & $0.27$ & Torus  \\
    \hline \hline
\end{tabular}
\label{table:models}
\end{table}

For each source morphology and composition, we place an initial coarse grid with ejecta masses ($M_{ej}/{\rm M}_\odot$) having values $0.001, 0.003, 0.01,0.03, 0.1$ and ejecta velocities ($v_{ej}/c$) having values $0.05, 0.15,0.3$ for each of the two components making our initial grid of $(5\times 3)^2 = 225$ simulations (see Refs. \cite{2017Natur.551...80K, 2019MNRAS.489L..91C, 2020Sci...370.1450D,2020PhRvD.101j3002K, dietrich2017modeling, 2019MNRAS.489L..91C} for a discussion on expected ejecta masses). The base grid [$(5\times 3)^2 = 225$] for our simulations is the same as previously studied in Ref. \cite{kilonova-lanl-WollaegerNewGrid2020} to which we add additional simulations doubling the grid space via the active learning technique discussed in Ref. \cite{Ristic22}. The iterative process targets subsequent simulation parameters for investigation, based on the estimated uncertainty of a (simplified) Gaussian Process estimate for the emitted radiation from the existing simulation set. By the end of the iterative process we accumulate 450, 449, and 449 simulations for the three morphology and composition combinations, respectively. The TSwind1 and TSwind2 model families each have one fewer simulation in their training libraries due to isolated instances of single simulation processing error. We limit the iterative learning process to 450 simulations because in our experience from previous study (Ref. \cite{Ristic22}) sufficiently accurate surrogate models can be obtained with a sample size this large.

We set the outflow velocity based on prior understanding of expected ejecta \cite{2017Natur.551...80K, nedora2021mapping, radice2018binary}. The ejecta velocity profile can alter the location of the photosphere for any given wavelength and thus affect the resultant LC. Specifically, the near-UV radiation peak time can be moved from an hour to a day postmerger by simply changing the velocity distribution. In this paper, however, we assume a homologous velocity profile for the ejecta, which, i.e., it scales as $r/t$, where $r$ is the distance from the center and $t$ is the time postmerger. We do not include a third component that could attribute to the early blue peak.

\subsection{Motivation for alternate morphologies}
Inferences made about ejecta properties of the event AT2017gfo, derived from the kilonova LCs \cite{Ristic22}, are not consistent with contemporaneous forward-model predictions for ejecta masses from GW inferences of the masses deduced for GW170817. Our kilonova analysis suggests that much larger wind velocities (and masses) will be required, inconsistent with the modeling assumptions usually adopted for polar winds. 

Recently, Nedora \textit{et al}. \cite{Nedora2019Spiral, 2021ApJ...906...98N} have proposed an alternative mechanism to generate disk outflows: a spiral-wave-driven wind which generally distributes over a large solid angle. For the purposes of our phenomenological kilonova inference, which focuses on the nature of outflows rather than their origin, such an outflow will have a different morphology than the polar ``peanut" wind and ``torus" dynamical morphology assumed for outflows. 
Further, \citet{Breschi2021AT2017gfo} have performed inference using simplified LCs extracted from systematically explored one-, two- and three-component models with several isotropic and anisotropic morphologies. They find a strong preference for anisotropic models with the inclusion of multiple components leading to largest evidence. However, their isotropic model (ISO-DV) makes even the dynamical components isotropic, whereas in this paper we choose an anisotrpic dynamical ejecta shape (torus). Additionally, \citet{2020ApJ...889..171K} obtain a strong similarity in the LCs for AT2017gfo and for a two-component model with a postmerger wind ejecta that is spherically symmetric.

The effect of morphologies of the ejecta was also studied extensively in the past \cite{2021ApJ...910..116K}, where morphologies were shown to have a greater impact on the LCs than the ejecta masses and velocities. The effect manifests heavily in the form of altering the peak luminosity and peak times.
We therefore assemble an expanded archive of actively-learned kilonova models, with more options for the wind-driven outflow morphology, to assess whether we can better reconcile our kilonova inference with forward models for the ejecta. To maximize the impact of altering wind morphology in our study, we adopt spherical and peanut shaped wind model. Other detailed simulations and the studies above all naturally produce a morphology much more similar to our preferred (torus) morphology.

\subsection{Simulation interpolation}
\label{sec:sim_interpoation}
Using the set of actively-learned simulations, we follow our previous approach \cite{Ristic22} to interpolate the resulting AB magnitudes versus simulation parameters $(m_{d},v_d,m_w,v_w)$, time postmerger, viewing angle, and wavelength bands. Interpolation over simulation parameters and for each band is performed by Gaussian process regression at fiducial reference times and fiducial angles ($0,30,45,60,75,90$ degrees). A continuous LC over time and angle follows by stitching together these fiducial results with simple low-dimensional interpolation.
Unless otherwise noted, we quantify the performance of our interpolation with the RMS difference between our prediction and the true value. Because of the substantial dynamic range of our many outputs, we interpolate in AB magnitudes using the LSST \textit{grizy} and 2MASS \textit{JHK} bands as our reference bands. Our raw LCs are calculated in absolute magnitudes, i.e. at a fiducial distance of 10 pc to the source.

\begin{figure}
    \includegraphics[width=0.5\textwidth]{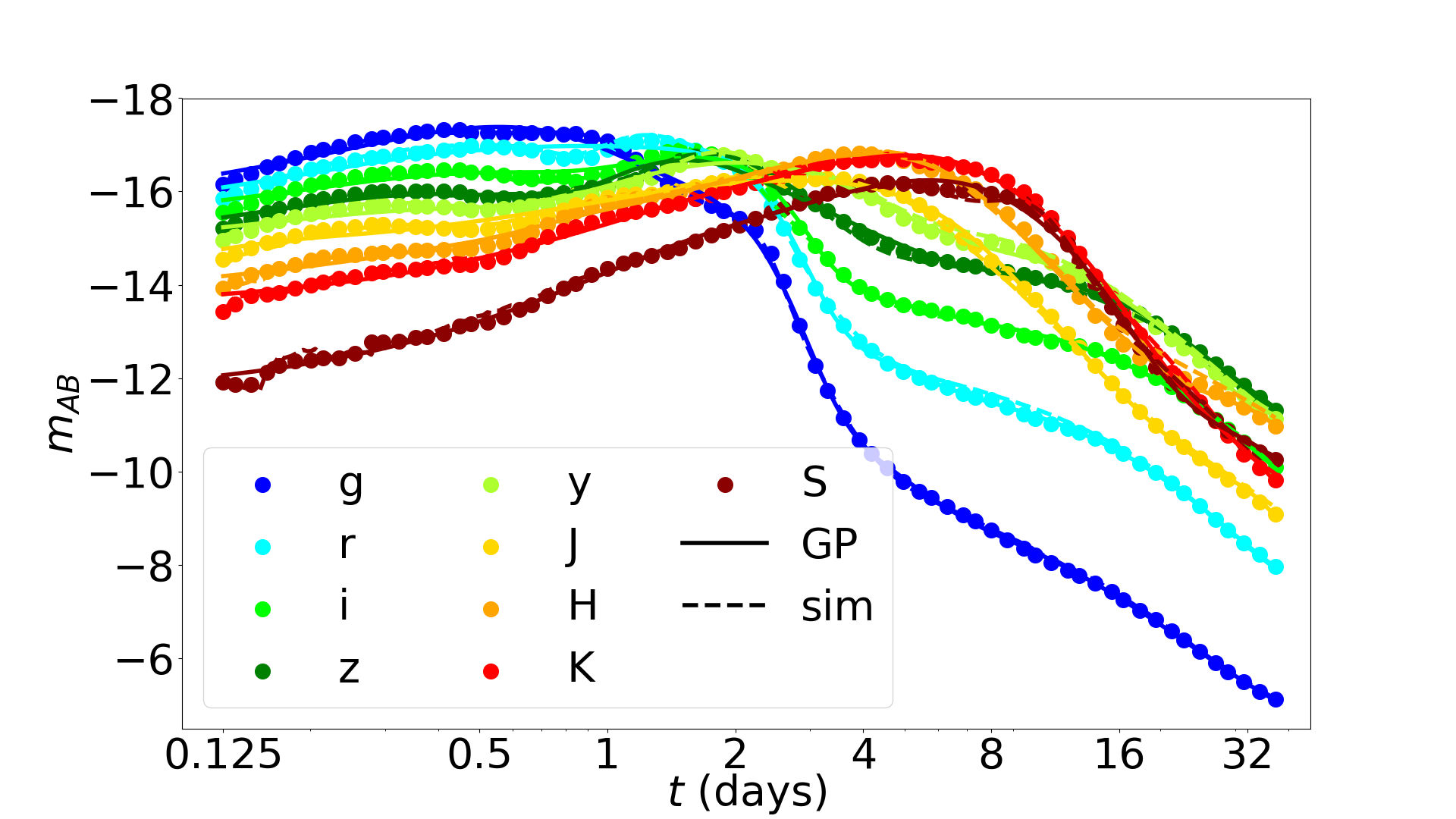} \\
    \includegraphics[width=0.43\textwidth]{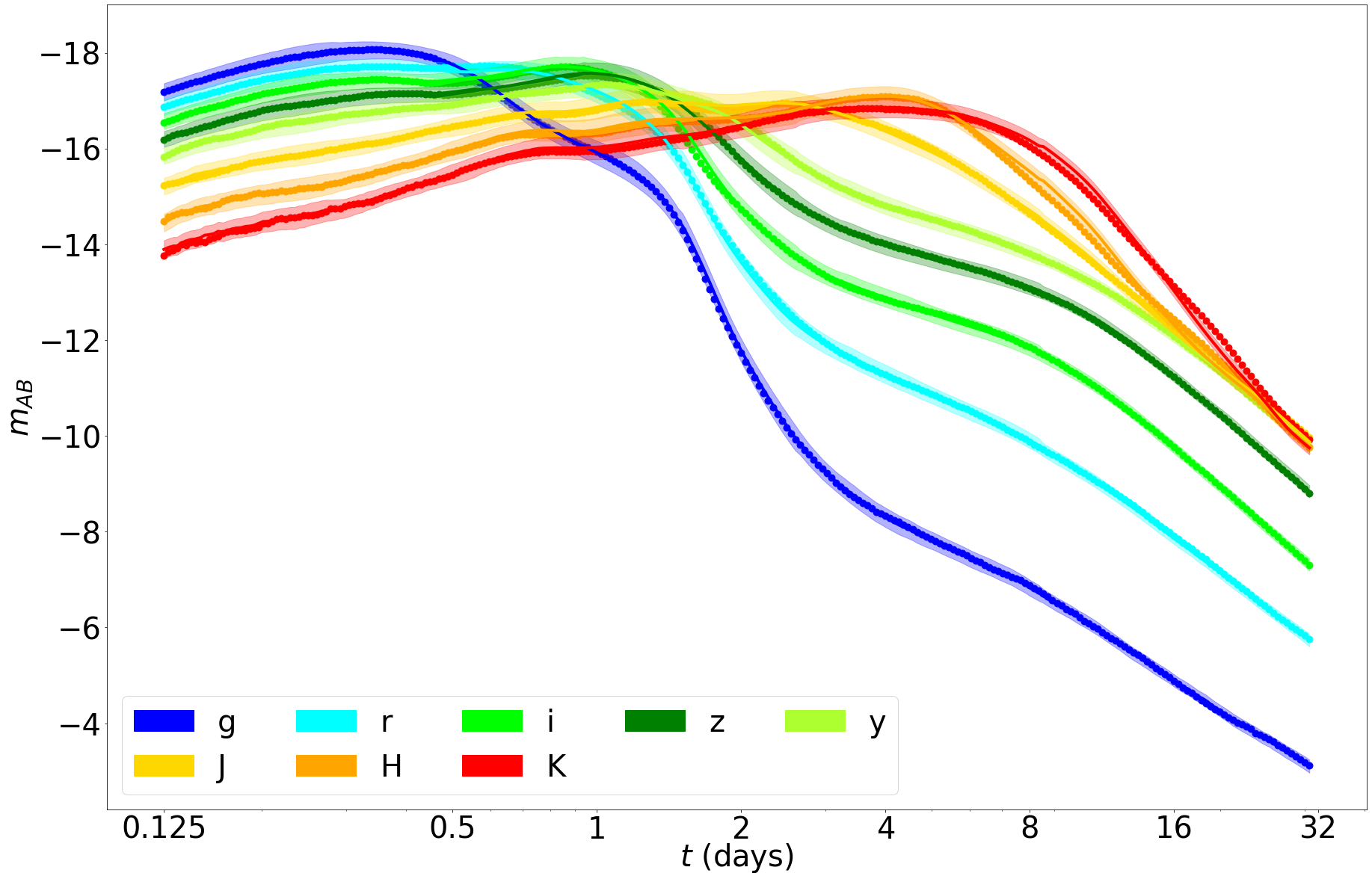}
    \caption{Illustrating off-sample interpolations for two morphologies. \emph{Top panel}: The bottom panel from Fig. 4 of Ref. \cite{Ristic22} 
    showing interpolation of various LCs versus time for the  TPwind2 morphology adopted in that paper. Different colors denote different filter bands, described in the legend. The dashed lines show full simulation output for each band. The colored points show our interpolated magnitude predictions at the evaluation times.
    The simulated parameters and viewing angle for this configuration are ($m_d/{\rm M}_\odot$, $v_d/c$, $m_w/{\rm M}_\odot$, $v_w/c$) = (0.097, 0.198, 0.084, 0.298) at $0^\circ$ \cite{Ristic22}. \emph{Bottom panel}: As above, but for the TSwind2 model morphology. The shaded region surrounding each solid curve shows our estimated GP fitting  uncertainty at each time. The differences between these two panels illustrate the impact of outflow morphology on our results.}
    \label{fig:morphology_effects}
\end{figure}

In Fig. \ref{fig:morphology_effects} we show our surrogate model predicting LCs for off-sample outflow. The prediction is for outflow parameters ($m_d/{\rm M}_\odot$, $v_d/c$, $m_w/{\rm M}_\odot$, $v_w/c$) = (0.097, 0.198, 0.084, 0.298). These parameters are chosen such that they are neither on the primary training grid nor the actively chosen simulation parameters. Top panel shows the parameter LC predicted for TPwind2 configuration, as was performed in \cite{Ristic22}, whereas the bottom panel shows the same for TSwind2 model. Hence, apparent differences in the two predictions arise solely due to differences in the wind morphology.

\begin{figure}
    \includegraphics[width=0.5\textwidth]{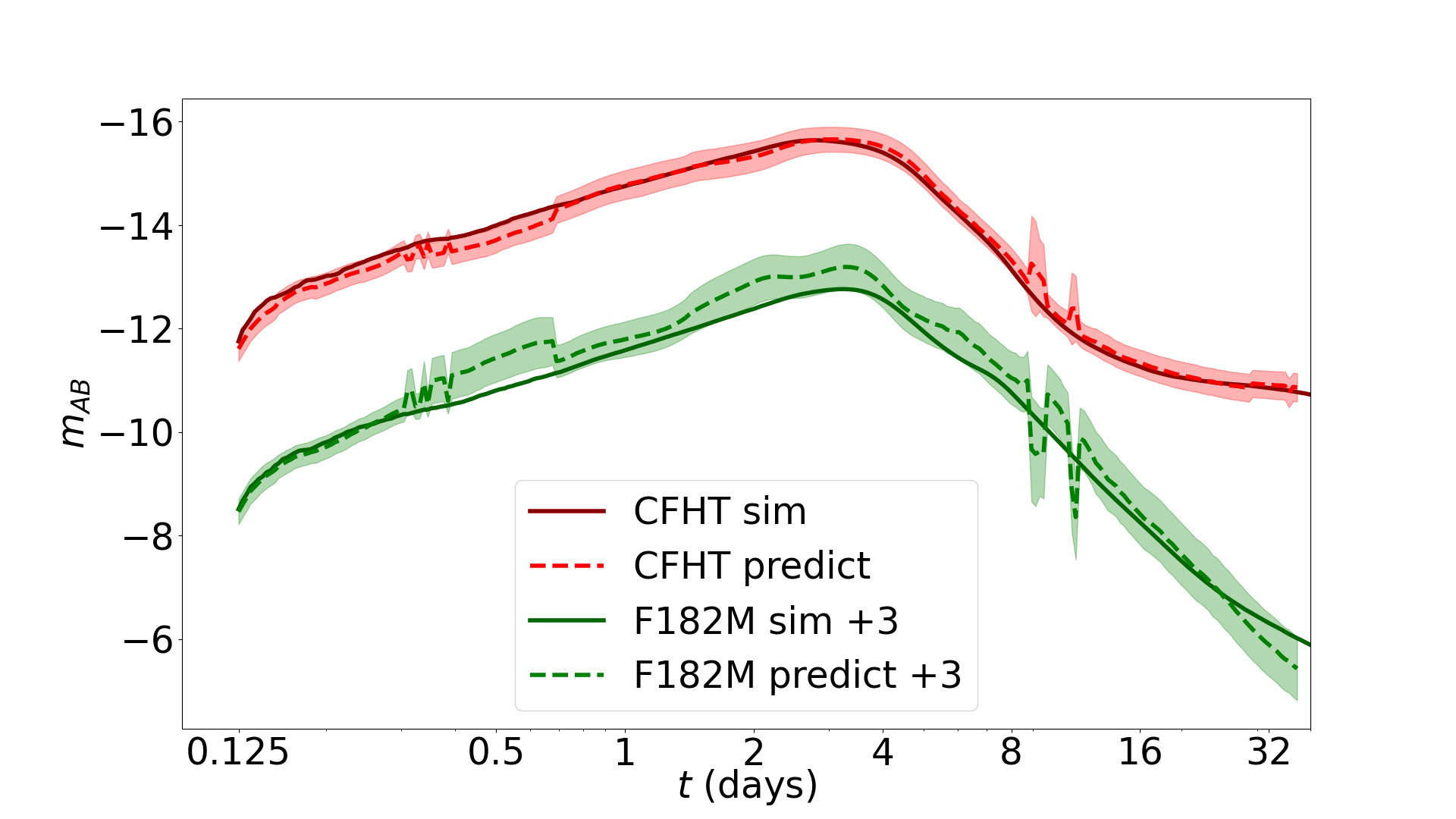}
    \caption{Wavelength interpolation using our surrogate model for an off-sample wavelength bands CFHT (1451 nm, i.e., between the $J$- and $H$-bands) and F182M (1839 nm, between $H$- and $K$-bands), generated for the TPwind1 model with parameters ($m_d/{\rm M}_\odot$, $v_d/c$, $m_w/{\rm M}_\odot$, $v_w/c$) = (0.014, 0.183, 0.085, 0.053) viewed along the symmetry axis $\theta=0$. The dashed curves (red for CFHT and green for F182M) along with their respective bands indicate the predicted LC from our surrogate model and solid curves indicate the LC generated from the code \texttt{SuperNu} via the mechanism explained in Sec. \ref{sec:simulation}.
    }
    \label{fig:wavelength}
\end{figure}

As noted earlier, our GP training scheme interpolates between filter bands, over a nominal \emph{wavelength} parameter. As a result, our GP approach can estimate LCs for filters both included and not included in our original training set. We verify the fidelity of our wavelength interpolation LC prediction by comparison to \texttt{SuperNu} spectral output convolved with two new filters. Fig. \ref{fig:wavelength} illustrates our wavelength interpolation LC prediction in dashed curves compared to the \texttt{SuperNu} spectral output in solid curves convolved with two distinct real filters not included in our original training set. One of these filters is the the WIRCam 8105 broadband filter \cite{2004SPIE.5492..978P} from the Canada-France-Hawaii Telescope (CFHT), part of the Mauna Kea Observatory, with an effective wavelength of $1.45\ \mu \textrm{m}$. This filter has minimal overlap with the $H$-band filter, thus serving as an independent test of wavelength interpolation ability. We also show results for the JWST/NIRCam.F182M filter \cite{2016jdox.rept......}, which lies between the $H$- and $K$-bands with small overlap with the $K$ band and has an effective wavelength of $1.84\ \mu \textrm{m}$.\footnote{These filter response functions were taken from the SVO Filter Profile Service \cite{2020sea..confE.182R}.}

Our GP models occasionally produce glitches as can be seen at $t~\sim 10$ days in Fig. \ref{fig:wavelength}. These glitches occur only for the wind1 composition models for both TS and TP morphologies (also visible in Fig. \ref{fig:lc}) and affect the quality of the predicted LC significantly at times beyond 4 days. This error-prone training may be resolvable by developing a sophisticated hyperparameter selection method that restricts the hyperparameters to the region of the preceding parameters hence avoiding large jumps.

\subsection{Ejecta parameter inference}
\label{sec:method_ejecta_parameter_inference}
\begin{figure*}
    \centering
        \includegraphics[width=0.9\columnwidth]{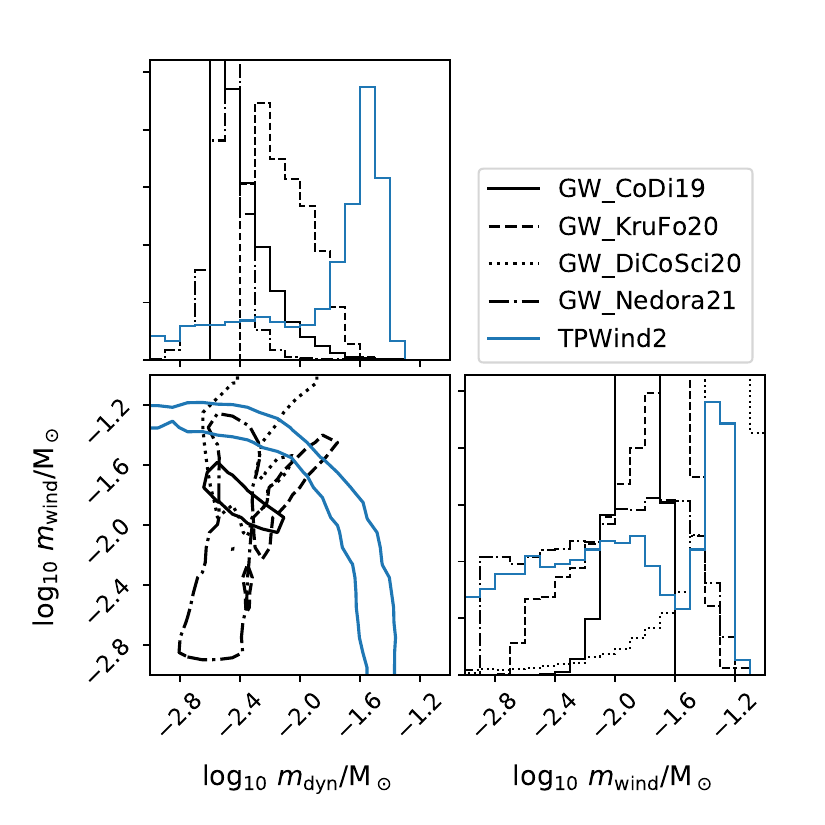}
        \includegraphics[width=1.1\columnwidth]{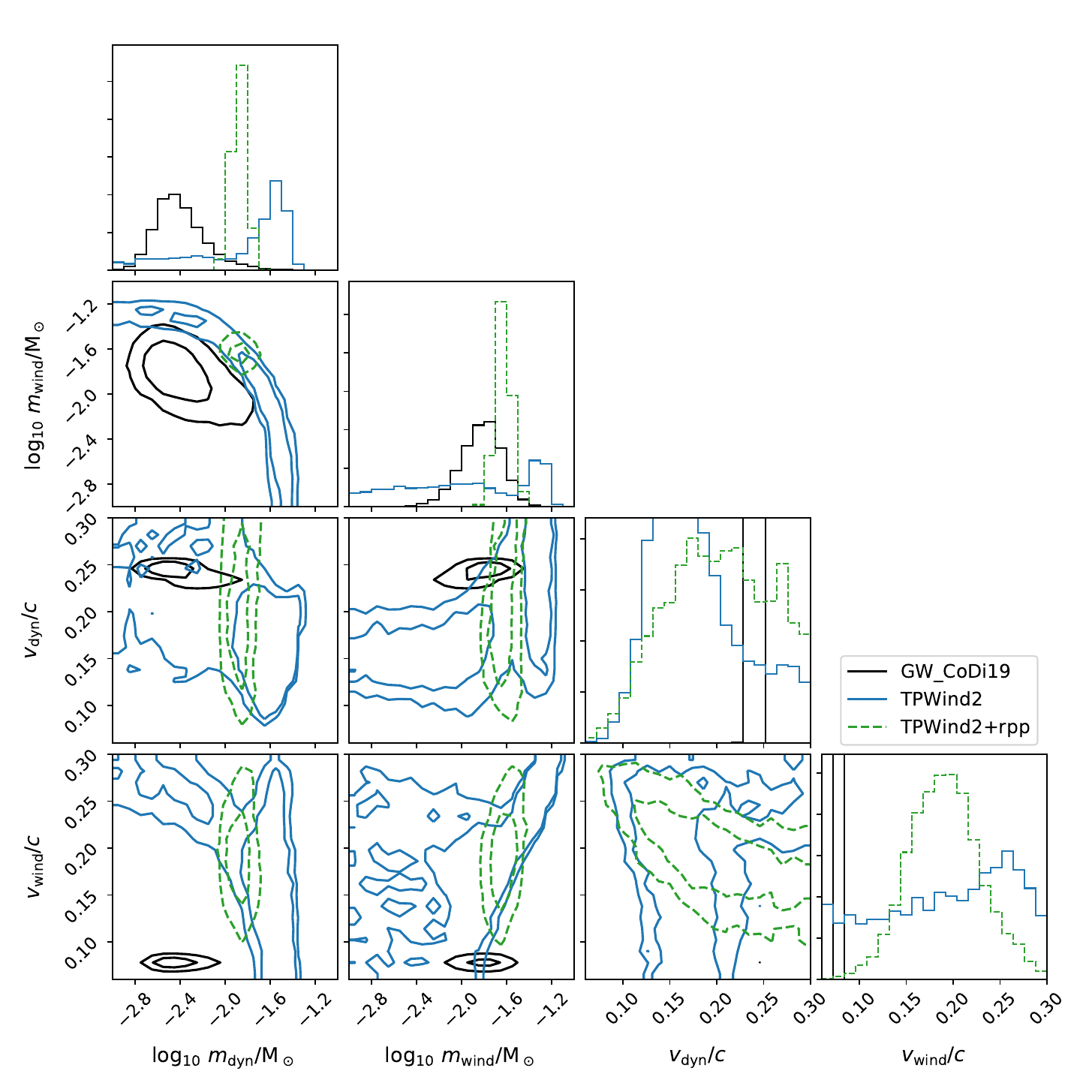}
    \caption{\emph{Left panel}: Illustration of systematics in modeling ejecta from binary mergers, compared to a fiducial inference of those ejecta parameters. Black curves are deduced from inferences on GW170817 mass estimates combined with different models for the ejecta; for example the black solid line is derived from Eqs. (\ref{equation:m_disk_CoDi19})--(\ref{equation:m_dyn}). To isolate the effect of ejecta modeling systematics, all predictions are shown under the optimistic assumption of a perfectly known nuclear equation of state, here APR4 \cite{akmal1998equation}, without adding additional uncertainties to account for estimated fitting errors in the relations between ejecta and binary parameters. While the ejecta predictions for GW170817 are relatively consistent with one another, they are in tension with our interpretation of the GW170817-associated kilonova's ejecta assuming a TPwind2 outflow form. Curves correspond to 90\% credible regions.
    \emph{Right panel}: Parameter inference for LC based on EM paper prior (uniform in velocities, \emph{log}-uniform in masses) \cite{Ristic22} (blue-solid) and \emph{r}-process prior \cite{2022arXiv220602273R} (green-dashed), but allowing for relative systematic uncertainty $\exp(\pm 0.3)$ in both ejecta masses. The inner and outer curves on the two-dimensional plots correspond to 68\% and 95\% credible regions.
    The ejecta estimate illustrated from GW includes no systematics, either from uncertainty in the EOS or from intrinsic uncertainty in the estimate itself; compare to Fig. \ref{fig:modified_pe_comp}. }
    \label{fig:modified_pe}
\end{figure*}

Following Ref. \cite{Ristic22} we use conventional Bayesian techniques to reconstruct the ejecta parameters $(m_d,v_d,m_w,v_w)$ and emission direction $\theta$ most consistent with GW170817, using the known source emission distance and taking into account prior information about the source orientation direction. As previously, we adopt uniform priors in velocity and $log$-uniform for masses over the limiting ranges. As a concrete example, the blue contours in Fig. \ref{fig:modified_pe} show previously-presented results from \cite{Ristic22} for two-component ejecta properties assuming the TPwind2 morphology.

While our default inference makes no additional assumptions about the ejecta, we separately also require that the merger from AT2017gfo produces a full \emph{r}-process element spectrum, consistent with the solar spectrum. Doing so constrains the ratio of $m_w/m_d$ to be close to 13.90 (1.76) for the wind1 (wind2) model assuming FRDM2012 as the nuclear mass model \cite{2016ADNDT.109....1M} and fission model by \citet{2010A&A...513A..61P} as described in \cite{2022arXiv220602273R}. The blue curves in the right panel of Fig. \ref{fig:modified_pe} show an analysis assuming this prior for the relative amounts of the two ejecta components.

The inferred ejecta properties with these Torus/Peanut models are, however, somewhat surprising in the context of conventional theoretical expectations about the two components. Most notably, our analysis requires a substantial wind velocity $\simeq 0.2 c$, consistent with previously-reported phenomenological estimates (see, e.g., Table 2 or Fig. 5 in \cite{2021ApJ...906...98N}), but in contrast to most theoretical expectations \cite{2020ARNPS..7013120R}.
This high wind velocity allows the peanut-morphology wind (which, in our calculations, explains the rapidly decaying blue component after its early peak) to have the short timescales required for the blue component, while simultaneously matching the target blue luminosity early on.
By contrast, the ``dynamical" red torus component can be consistent with GW170817's associated emission with a wide range of velocities \cite{Breschi2021AT2017gfo, 2020ARNPS..7013120R}. In earlier studies \cite{2021ApJ...910..116K}, it was noted that the dynamical ejecta mass was poorly constrained, due to the lanthanide curtaining effect \cite{2015MNRAS.450.1777K}. Since lanthanide curtaining occurs in models with greater velocity difference in the dynamical and wind components, this makes viewing angle dependence (caused by the curtaining) more velocity dependent than mass dependent \cite{2021ApJ...910..116K}.

\subsection{Contemporary forward models for neutron star merger ejecta}
\label{sec:mass_velocity_models}

After using parameter inference to deduce the possible immediate outflow responsible for the kilonova associated with GW170817, we arrive at inferred ejecta masses and velocities for that outflow. For context and to help guide our investigations, we compare these inferred ejecta properties with the predictions provided by selected contemporary forward models, which attempt to estimate the ejecta properties from inferred binary properties \cite{2019MNRAS.489L..91C, 2020Sci...370.1450D, Pang2022NMMA,2020PhRvD.101j3002K, dietrich2017modeling, 2019MNRAS.489L..91C}. We use four discrete models for this purpose, summarized in Table \ref{table:nr_models}, and present details about them here.

We employ the following estimates for the dynamical and wind ejecta mass \cite{2019MNRAS.489L..91C}. First, the disk mass is estimated as their Eq. (1),
\begin{align}
    \log_{10}(m_{\rm disk}[M_{\rm tot}/M_{\rm thr}]) & \nonumber \\
    = \max \bigg( -3, ~ a ~\bigg( &1 + b \tanh \bigg[ \frac{c - M_{tot}/M_{thr}}{d}\bigg]\bigg) \bigg)
    \label{equation:m_disk_CoDi19}
\end{align}
with $a=-31.335$, $b=-0.9760$, $c=1.0474$, $d=0.05957$. $M_{\rm tot}$ is the sum of neutron star masses, and $M_{\rm thr}$ is the threshold mass for the binary to undergo a prompt collapse to a BH after merger, which can be estimated by (e.g., \citet{bauswein2013prompt})
\begin{align}
    M_{th}=\bigg(2.38-3.606 \frac{M_{TOV}}{R_{1.6}}\bigg) M_{TOV},
\end{align}
where $M_{TOV}$ is the maximum gravitational mass of a stable nonrotating neutron star given the EOS and $R_{1.6}$ is the radius of the neutron star for the EOS with a mass 1.6 $\rm{M}_\odot$. The estimated uncertainty on this relation is of roughly a factor of 2, or 0.02 ${\rm M}_\odot$. The wind ejecta mass $m_{\rm wind}=\xi m_{\rm disk}$ where $\xi=0.3$, with an estimated uncertainty of $O(1)$. The dynamical ejecta mass is estimated by their Eq. (2),
\begin{align}
    \log_{10} m_{\rm dyn} =  \left[ a\frac{(1-2C_1)m_1}{C_1} + b ~ m_2 \bigg(\frac{m_1}{m_2} \bigg)^n + \frac{d}{2} \right]
    + [1\leftrightarrow 2],
    \label{equation:m_dyn}
\end{align}
where in this expression $m_1$ and $m_2$ are the neutrons star masses, $a=-0.0719$, $b=0.2116$, $d=-2.42$, $n=-2.905$, and $C_i$ are the compactnesses of the two neutron stars. $[1\leftrightarrow 2]$ is a shorthand for the preceding terms in the expression with permuted subscripts. The estimated uncertainty is for this relation is $7\times 10^{-3} {\rm M}_\odot$ when calculated linearly or up to 36\% when calculating for $\log_{10}m_{dyn}$. This model is labeled ``CoDi19'' in the relevant plots.

For a more recent estimate, we also utilize the updated model from Ref. \cite{2020Sci...370.1450D} that has been calibrated for a wider range of binary mass ratios. A new disk mass fitting is calculated in this study and is given in Eqs. (S4)--(S6) in Ref. \cite{2020Sci...370.1450D} and is the same as Eq. (\ref{equation:m_disk_CoDi19}) with updated constants. The fitting parameters here are $a=a_0 + \delta a \cdot \xi $, $b= b_0 + \delta b \cdot \xi $, and the resulting constants are $\xi= \frac{1}{2}\tanh (\beta (\hat{q} - \hat{q}_{trans})) $, $a_0= -1.5815, ~\delta a = -2.439, ~b_0 = -0.538, ~\delta b = -0.406, ~c = 0.953, ~d = 0.0417, ~\beta = 3.910, ~\hat{q}_{trans} = 0.900$. This model is labeled ``DiCoSci20'' in the relevant plots.

As a complementary estimate to highlight their systematic error, we also provide postprocessing estimates provided with an alternative set of fits from Kruger and Foucart \cite{2020PhRvD.101j3002K}. In this approach, the disk mass is estimated as
\begin{align}
    m_{disk,KF}= m_1 \text{max}[5\times 10^{-4}, (a C_1+c)^d],
\end{align}
where $C_1$ is the compactness of the lighter of the two neutron stars, $a=-8.1324$, $c=1.4820$, $d=1.7784$ with an error of order 40\%. The dynamical mass is estimated as
\begin{align}
\frac{m_{dyn,KF}}{10^{-3} {\rm M}_\odot}&=
    \left(\frac{a}{C_1}+b \left(\frac{m_2}{m_1}\right)^n + c~ C_1 \right) m_1 + [1\leftrightarrow 2],
\end{align}
where $a=-9.3335$, $b=114.17$, $c=-337.56$, and $n=1.5465$, where negative values imply zero ejecta. This model is labeled ``KruFo20'' in the relevant plots.

For broader context we also report estimates for the dynamical ejecta mass from GW170817 using the more inclusive and most recent fitting method by Ref. \cite{nedora2021mapping}. Among the available choices of fitting, we use the quadratic fitting of the form 
\begin{equation}
    \log_{10} (m_{dyn}/{\rm M}_\odot) = b_0 + b_1 q + b_2 \tilde \Lambda + b_3 q^2 + b_4 q\tilde \Lambda + b_5 \tilde \Lambda^2,
\end{equation}
where $q$ is the mass ratio ($m_2$/$m_1$ mass), and $\tilde \Lambda$ is the reduced tidal deformability. Among the new models introduced in that paper, this model provides the greatest flexibility with both mass ratio and tidal deformability dependence, although we note the conspicuous absence of dependence on binary total mass. From this paper, we adopt the fitting coefficients obtained for the reference set + M0/M1 set (labeled as M0RefSet and M0/M1Set), which include the best available physics, specifically excluding analyses that omit neutrino reabsorption and have pertinent systematic differences with the reference calibration adopted here. The parameters thus used are (from Table 4 of Ref. \cite{nedora2021mapping}):
$b_0 = -1.32$, $b_1 = -3.82 \times 10^{-1}$, $b_2 = -4.47 \times 10^{-3}$, $b_3 = -3.39 \times 10^{-1}$, $b_4 = 3.21 \times 10^{-3}$, $b_5 = 4.31 \times 10^{-7}$. We exclude a disk mass model from Ref. \cite{nedora2021mapping} due to large  systematics and inconsistent definition between the original datasets. Instead we employ the fit by \cite{2019MNRAS.489L..91C}. This model is labelled ``Nedora21'' in the relevant plots.

For ejecta velocities in each model we choose the results from Ref. \cite{dietrich2017modeling, 2019MNRAS.489L..91C}. Hence, for the dynamical ejecta we adopt the formula
\begin{align}
    v_{d}/c&=  a ~ \bigg(\frac{m_1}{m_2} \bigg)(1+c ~ C_1) + a ~ \bigg(\frac{m_2}{m_1} \bigg)(1+c ~ C_2)+b
    \label{equation:v_dyn}
\end{align}
with $a=-0.309$, $b=0.657$, and $c=-1.879$
with an estimated relative uncertainty of $20\%$ \cite{2019MNRAS.489L..91C}.
For the velocity of wind ejecta, we adopt
0.08c as was adopted by \cite{radice2018binary},
\begin{align}
    v_{w}&= 0.08 c ~.
    \label{equation:v_wind}
\end{align}

The solid-black contours in Fig. \ref{fig:modified_pe} show our estimated ejecta masses using these forward models for ejecta. As inputs to these expressions, we use the results of detailed parameter inference applied to GW170817, as described in Lange \textit{et al.} (in preparation); see also \cite{PhysRevD.107.024040}. Even allowing for reasonable uncertainties in these ejecta formulas, the tension with our inferred ejecta parameters is apparent: these models predict roughly $2\times$ smaller ejecta masses for both components, much slower wind velocities, and an extremely narrow range of expected dynamical ejecta velocities.

\begin{table}[tbpt]
\caption{\label{table:nr_models} Numerical relativity-based forward models employed in this work for GW170817 ejecta PE. The velocity fits are common for all. Dynamical ejecta velocity fit is given in Eq. (\ref{equation:v_dyn}) based on \cite{dietrich2017modeling, 2019MNRAS.489L..91C}. Wind velocity is set fixed at $v_w = 0.08c$ \cite{radice2018binary}.}
\begin{tabular}{@{\extracolsep{20pt}} l c c c c@{}}
    \hline \hline
    Name & $m_d$ & $m_w$ & Reference \\ \hline
    CoDi19      & \checkmark    & \checkmark    & \cite{2019MNRAS.489L..91C} \\
    DiCoSci20   & CoDi19        & \checkmark    & \cite{2020Sci...370.1450D} \\
    KruFo20     & \checkmark    & \checkmark    & \cite{2020PhRvD.101j3002K} \\
    Nedora21    & \checkmark    & CoDi19        & \cite{nedora2021mapping} \\
    \hline \hline
\end{tabular}
\end{table}

\section{Results}
\label{sec:results}

\subsection{Impact of morphology on interpolated light curves}
\label{sec:results_morphology}
Using simplified \cite{2017ApJ...850L..37P, Breschi2021AT2017gfo, Nicholl21} and more realistic \cite{2018MNRAS.478.3298W, bulla2019possis, 2021ApJ...910..116K, 2020ApJ...889..171K, kawaguchi2021low} models for kilonovae and the associated radiative transfer, several previous studies have demonstrated that the angular distribution of the outflow can imprint its signature on the outgoing radiation.
The first and second columns of Fig. \ref{fig:magnitude_versus_angle} illustrate the impact of outflow morphology on the interpolated LCs. The plots show $g$-, $y$-, and $K$-band luminosity as time proceeds along viewing angles for a sample model parameter ($m_d/{\rm M}_\odot$, $v_d/c$, $m_w/{\rm M}_\odot$, $v_w/c$) = (0.097, 0.198, 0.084, 0.298). As anticipated from the aforementioned prior paper, we notice a strong impact from the choice of morphology. In particular, relative to our polar outflow illustrated in the first column (TPwind1), the spherical wind outflow in the second column (TSwind1) produces relatively less angular variability of the LC, particularly in the bluer bands. 

A noticeable feature of TPwind1 LCs is that their angular dependence flips over in the late times, as can be seen by the apparent opposite curvature at $t = 4$ days and $t = 8$ days. This behavior is resulting from the dispersion of the dynamical torus component at late times. Compared to this result, the TS late time LCs are angle independent both for low and high electron fraction.

The TPwind1 model is able to also exhibits a peak in the blue band in the early times ($t = 0.5$ days) as is noticeable in the top-left panel. A change in the morphology to TSwind1 changes this blue band LC to be monotonically diminishing, noticeable in the top-middle panel. This will again be seen in a Fig. \ref{fig:lc} below where parameters fitting AT2017gfo with TPwind1 and \emph{r} process prior will produce the short-time blue kilonova peak. This can, however, be altered and other morphologies be made to produce the blue peak at desired times by fine-tuning component parameters, thus highlighting the challenges in replicating this observation.

The late time $K$ band is rather similar for all morphologies and compositions, and most of the differences are present in the earlier times. This is because, at late times, the photosphere has moved inwards substantially so morphology effects are minimized and the temperature at the photosphere is near the $K$ band.

\subsection{Impact of composition on interpolated light curves}
\begin{figure*}
    \centering
    \begin{subfigure}[b]{0.33\textwidth}
        \includegraphics[width=1.08\textwidth]{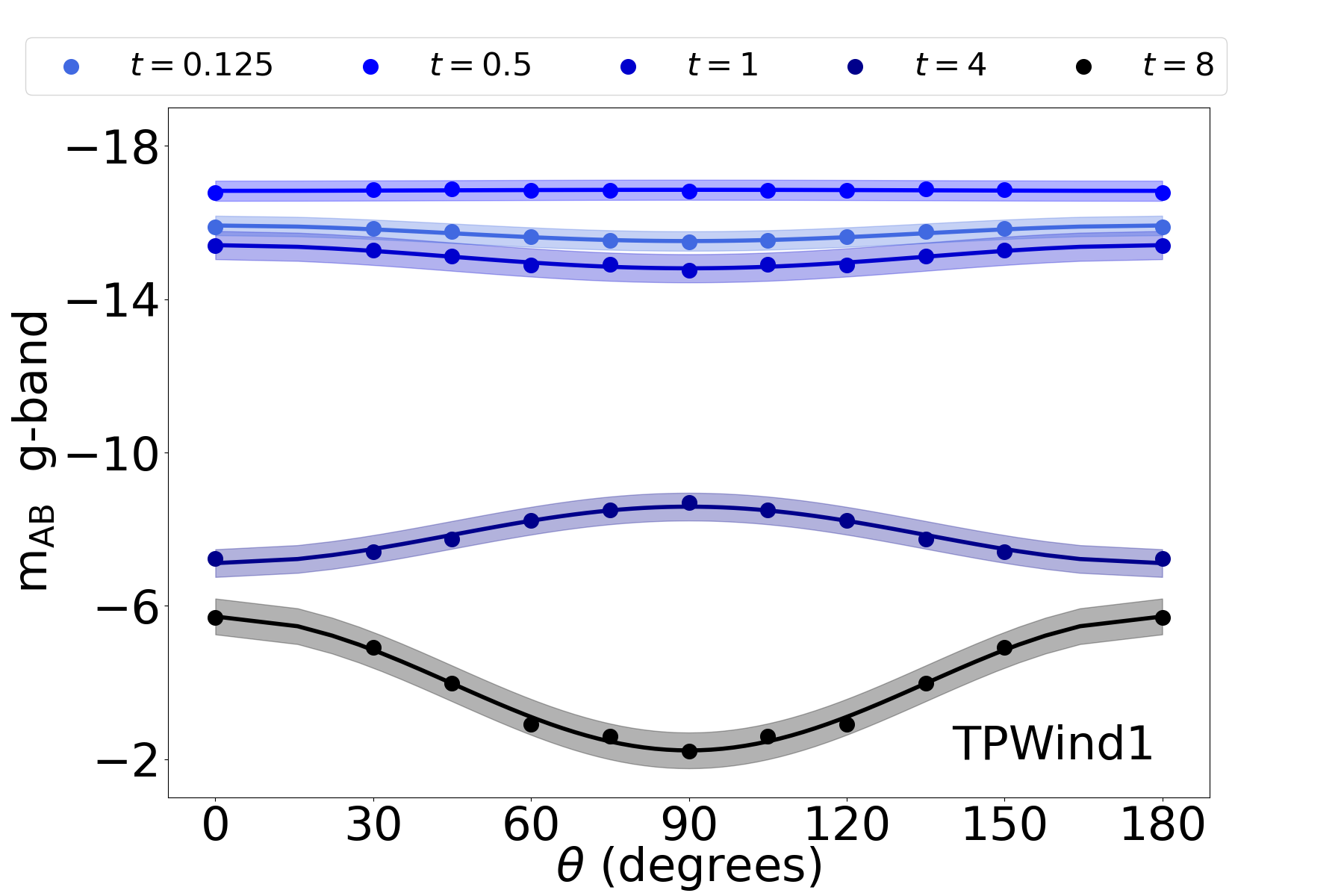}
    \end{subfigure}%
    \begin{subfigure}[b]{0.33\textwidth}
        \includegraphics[width=1.08\textwidth]{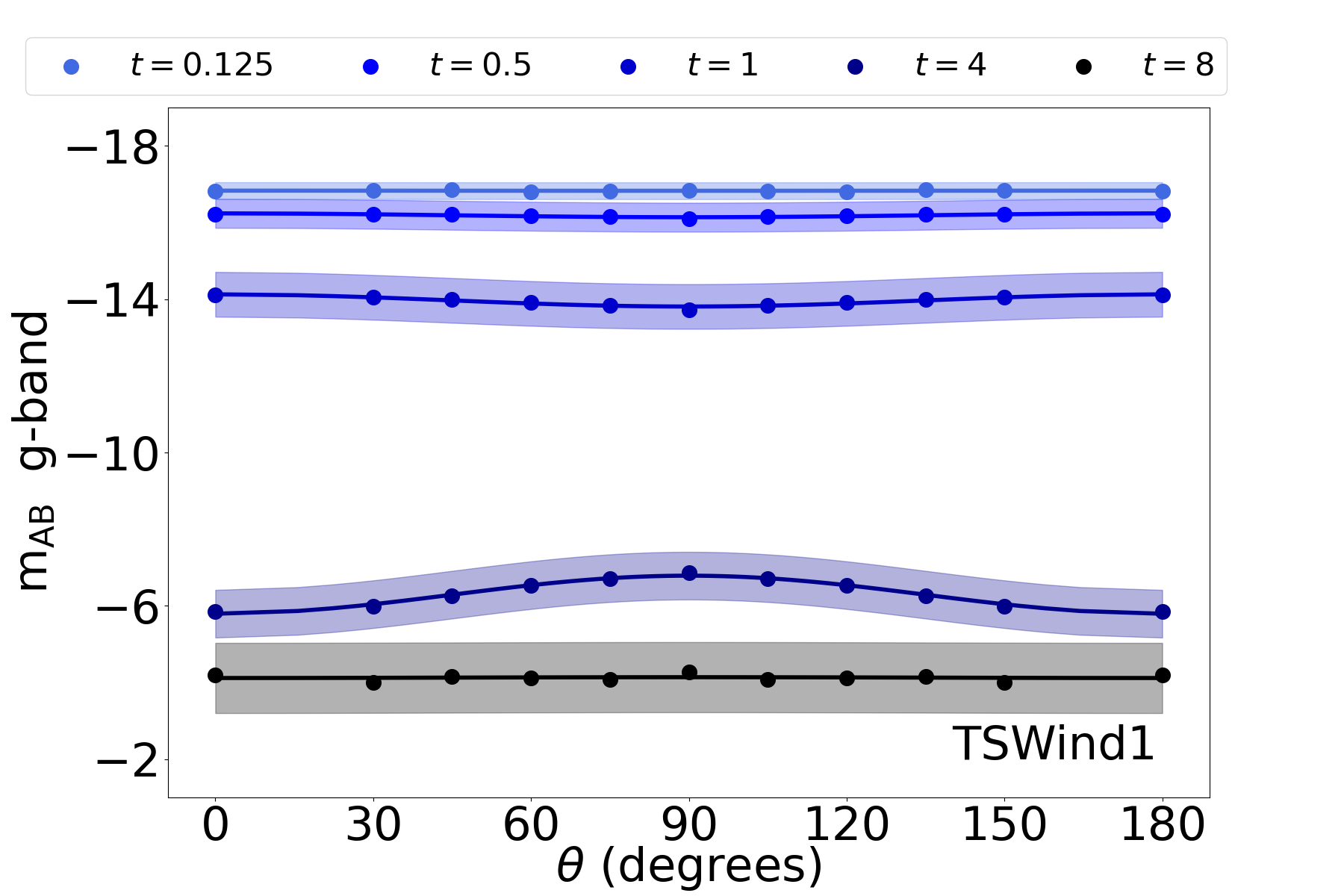}
    \end{subfigure}%
    \begin{subfigure}[b]{0.33\textwidth}
        \includegraphics[width=1.08\textwidth]{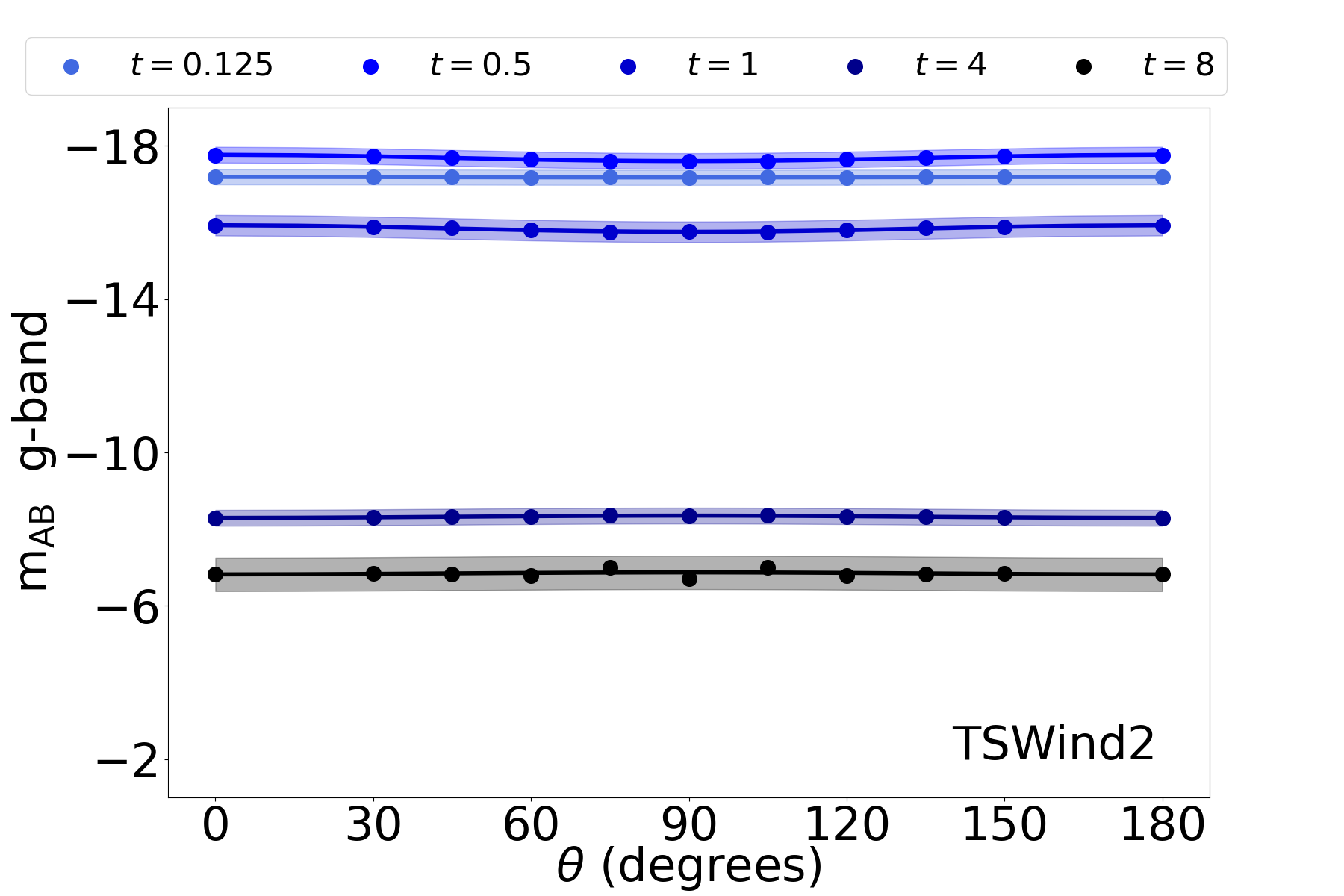}
    \end{subfigure}%
    
    \begin{subfigure}[b]{0.33\textwidth}
        \includegraphics[width=1.08\textwidth]{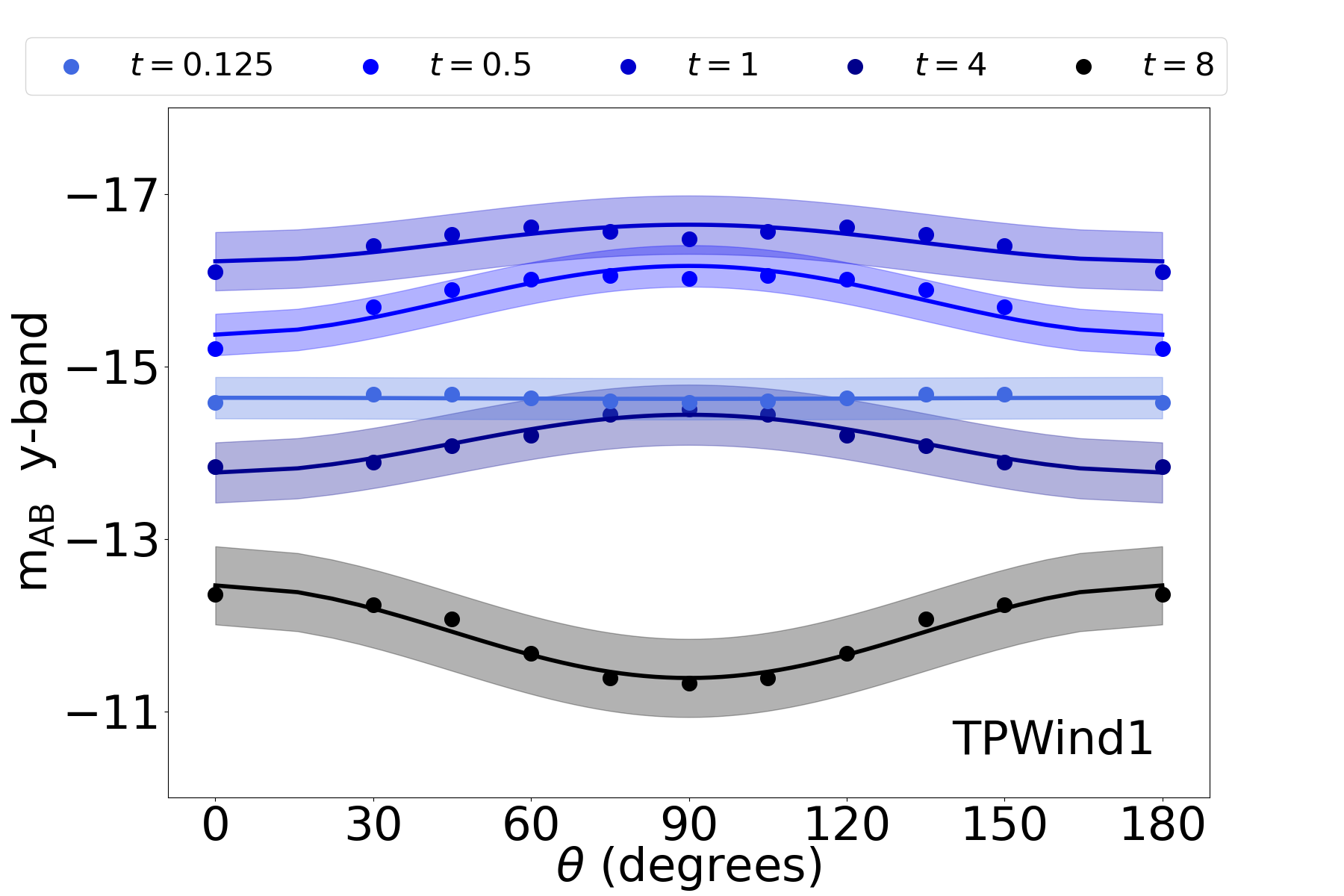}
    \end{subfigure}%
    \begin{subfigure}[b]{0.33\textwidth}
        \includegraphics[width=1.08\textwidth]{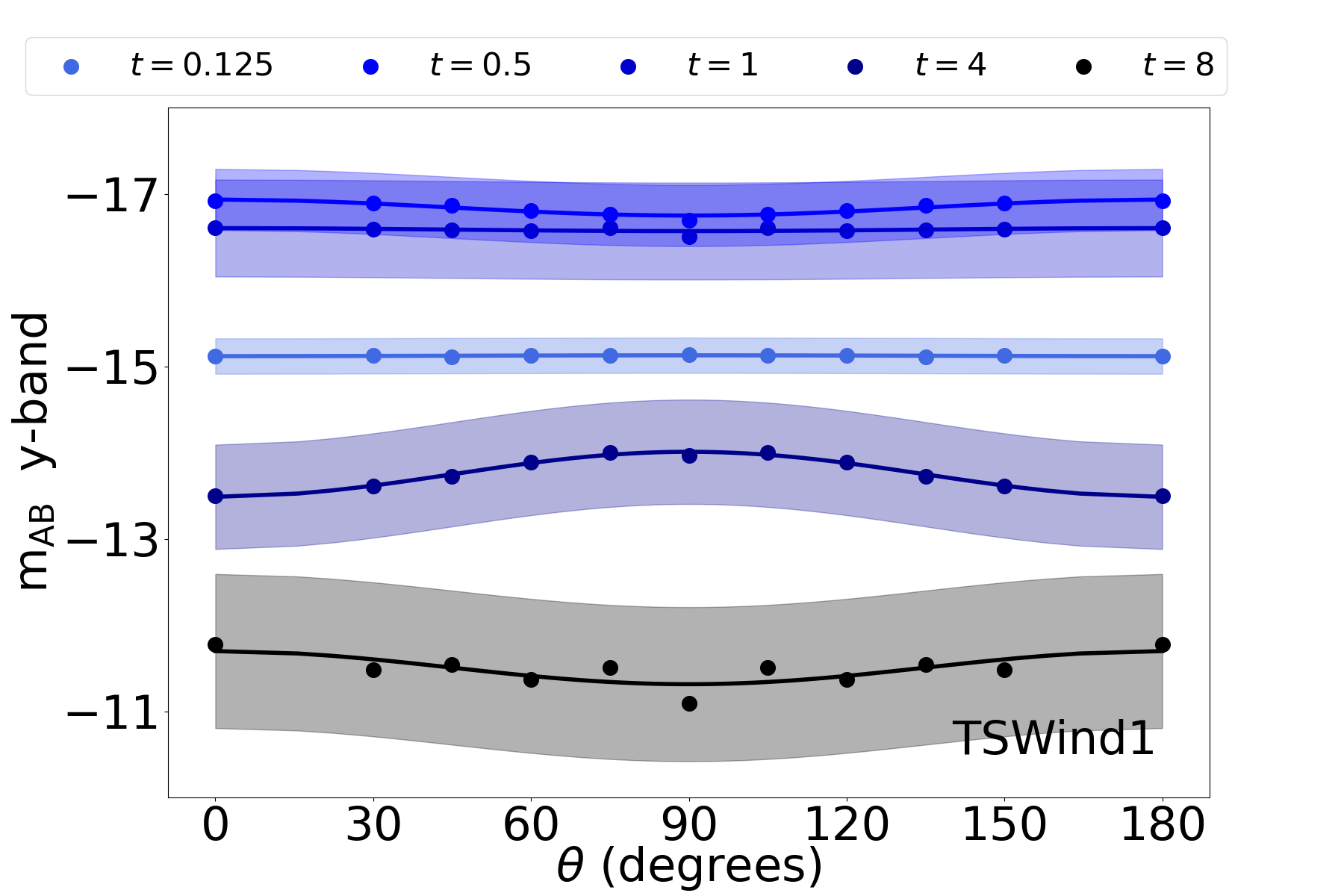}
    \end{subfigure}%
    \begin{subfigure}[b]{0.33\textwidth}
        \includegraphics[width=1.08\textwidth]{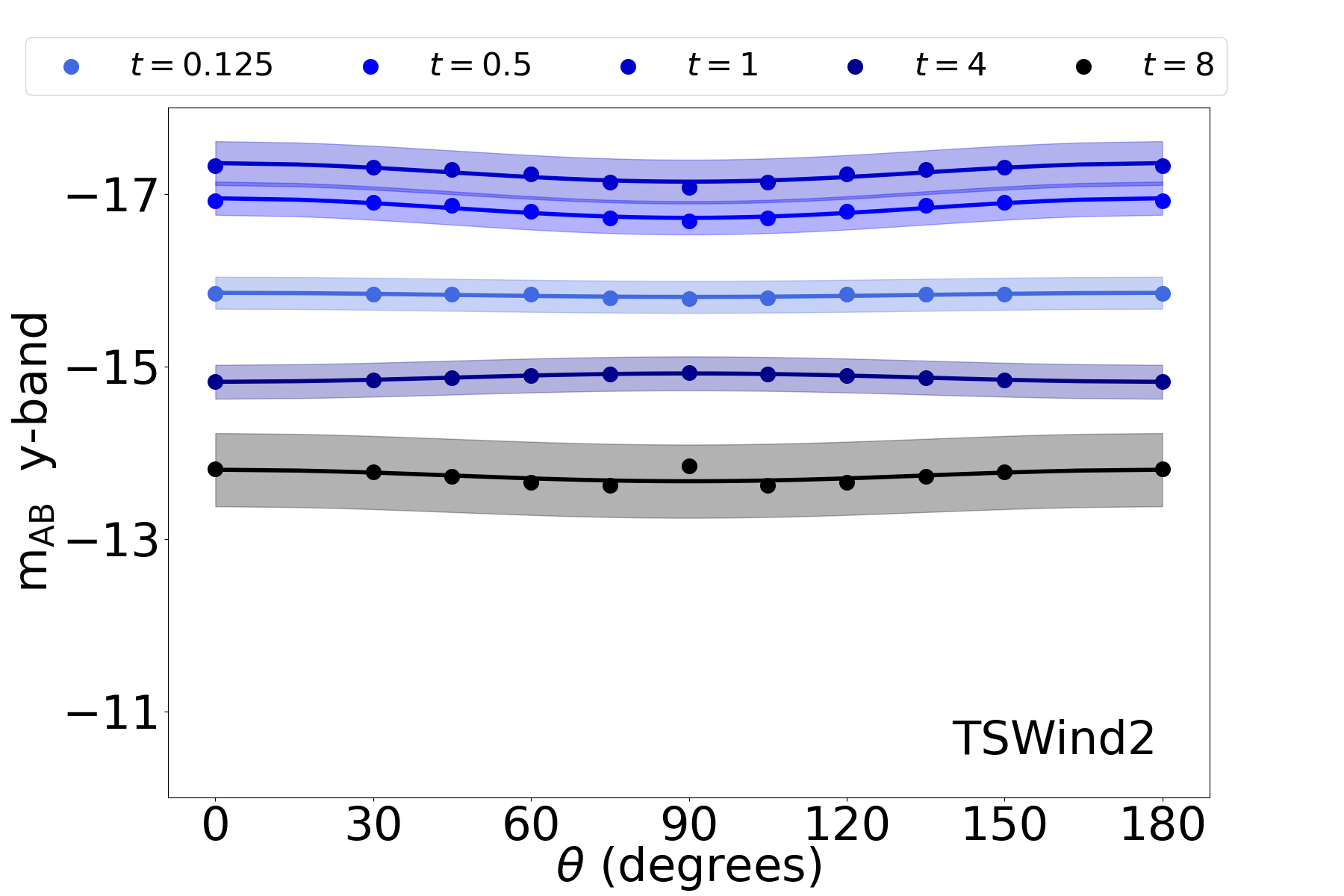}
    \end{subfigure}
    
    \begin{subfigure}[b]{0.33\textwidth}
        \includegraphics[width=1.08\textwidth]{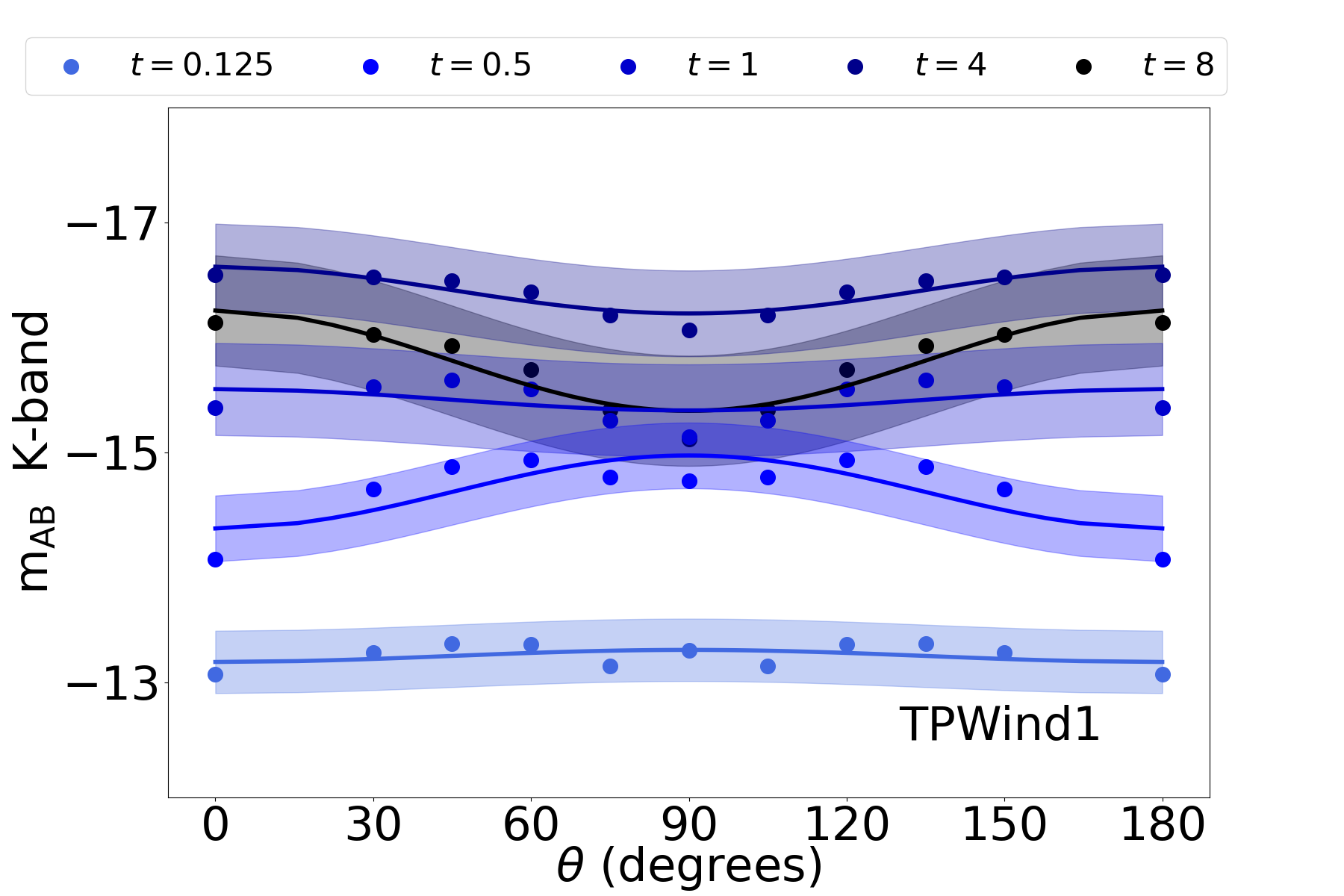}
    \end{subfigure}%
    \begin{subfigure}[b]{0.33\textwidth}
        \includegraphics[width=1.08\textwidth]{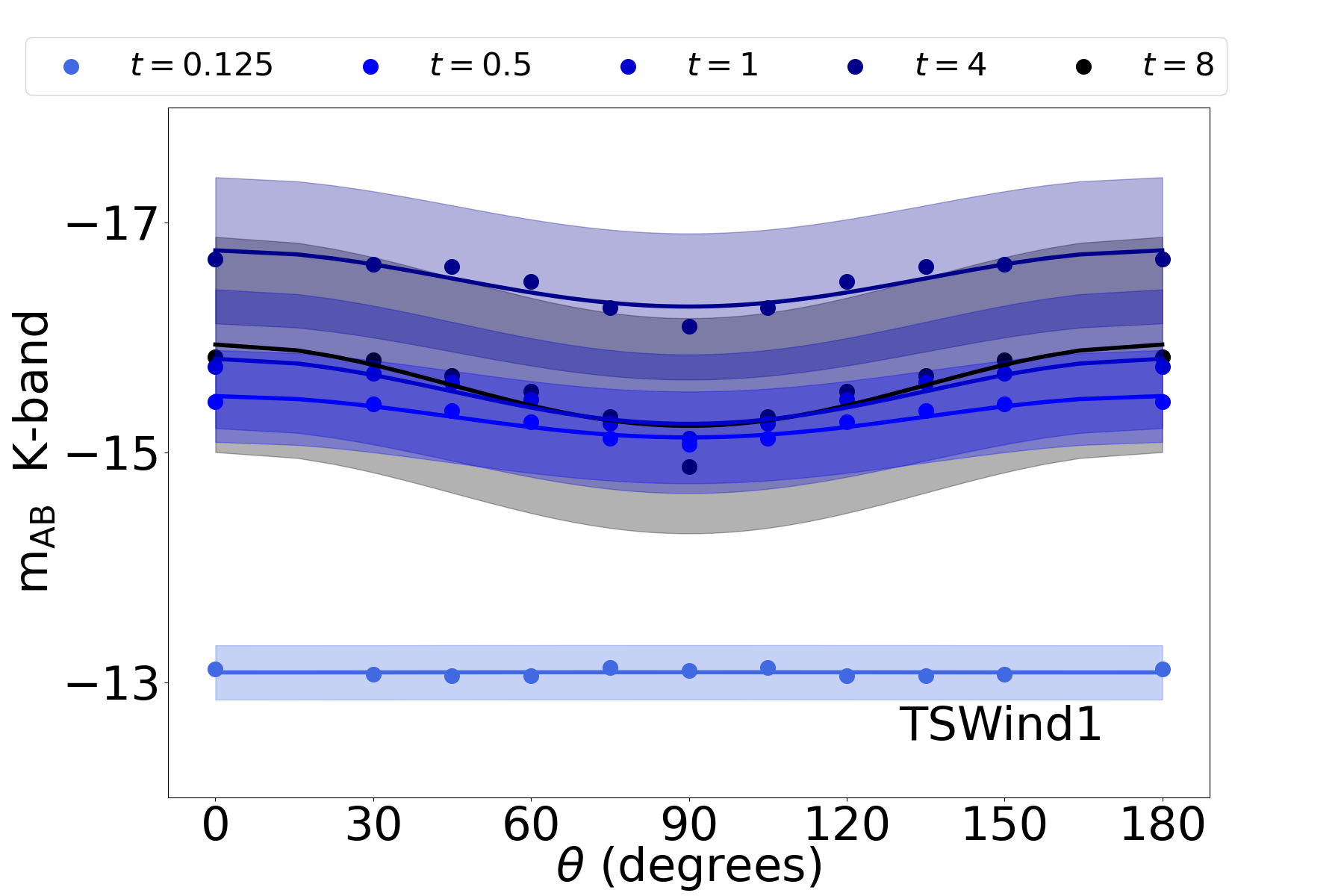}
    \end{subfigure}%
    \begin{subfigure}[b]{0.33\textwidth}
        \includegraphics[width=1.08\textwidth]{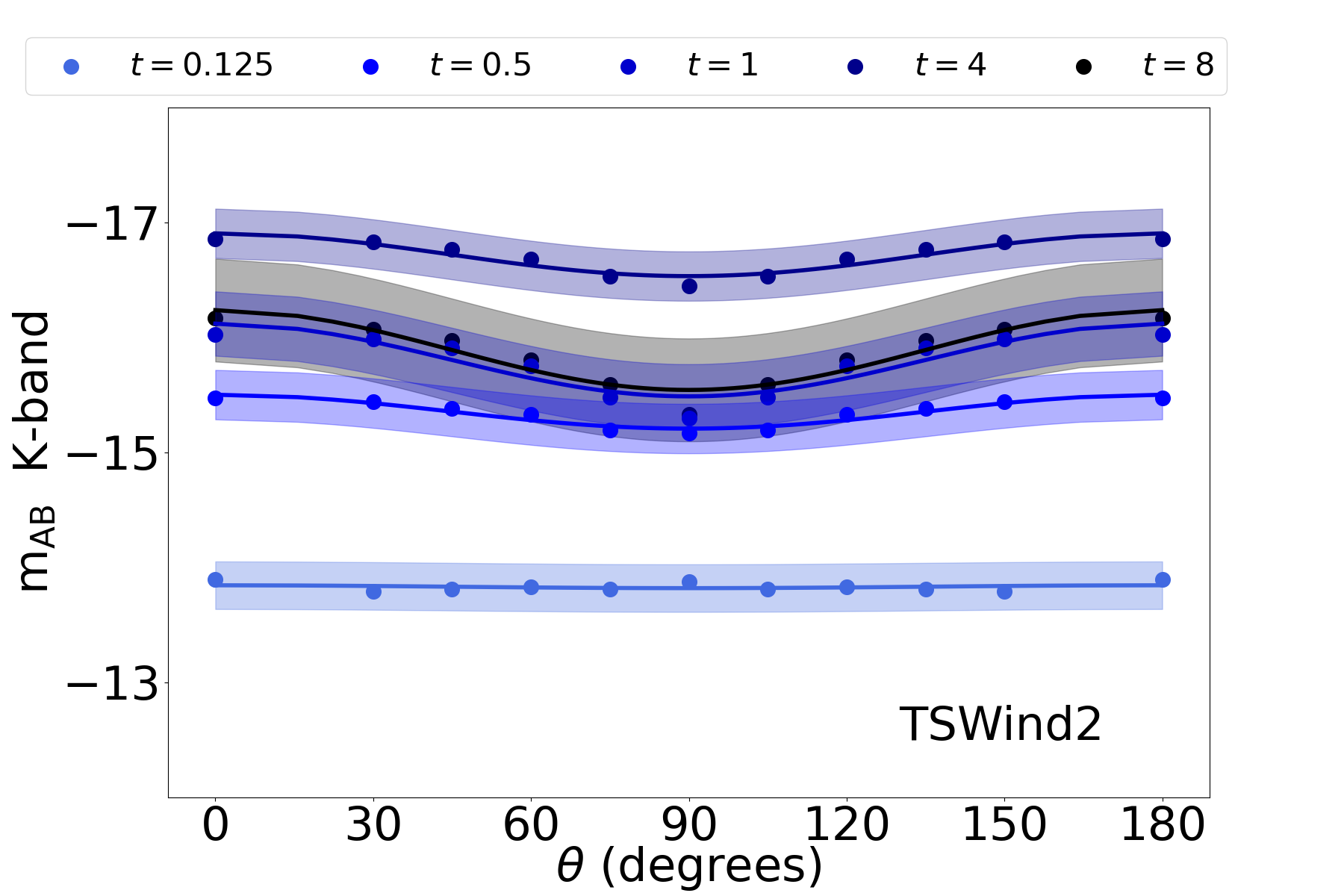}
    \end{subfigure}%
    \caption{Angular dependence for the three surrogate models for each morphology and composition combination, i.e., TPwind1, TSwind1, and TSwind2. This figure compares the $g$-, $y$-, and $K$-band luminosity (top to bottom) at select times as a function of viewing angle. Different colors indicate different extraction times. Points show simulation AB magnitude results vs angle $\theta$, while the solid curves and shaded regions show our prediction and its expected (statistical interpolation) uncertainty. All simulations and surrogate LCs use the parameters ($m_d/{\rm M}_\odot$, $v_d/c$, $m_w/{\rm M}_\odot$, $v_w/c$) = (0.097, 0.198, 0.084, 0.298); because this configuration has $v_w>v_d$ (i.e., wind outside the dynamical ejecta), the wind ejecta's emission and angular dependence could dominate trends versus angle. \emph{Left column}: Angular dependence for TPwind1 light bands. This model shows the largest viewing angle dependence. \emph{Center column}: Angular dependence for TSwind1. A change in angular dependence is highly apparent at late times particularly in the $g$ band. \emph{Right column}: Angular dependence for TSwind2. Differences from the first two models are apparent. This configuration shows the weakest angular dependence in its emission and the brightest emission.}
    \label{fig:magnitude_versus_angle}
\end{figure*}

The second and third columns of Fig. \ref{fig:magnitude_versus_angle} illuminate how ejecta composition ($Y_e$) impacts emitted radiation. As expected, we find brighter LCs at early and late times for lower $Y_e$ (i.e., wind2 is brighter than wind1) in the $g$-, $y$-, and other lower wavelength bands, and somewhat in the $K$ bands.
As a result, the choice of wind composition $(Y_e)$ is partially degenerate with wind ejecta mass $m_w$, which is not surprising since products like $Y_e m_{w}$ enter naturally into the initial conditions.

Further, their is almost no angular dependence in the TSwind2 and a small angular dependence in TSwind1 because the wind outflow surpasses the dynamical and dominates the kilonova producing a viewing angle-independent signal.
At late times the wind ejecta has substantially outflown the dynamical ejecta due to having $v_d > v_w$ and directs the emission. The TSwind2 model, due to its lower $Y_e$, has a higher wind-heating rates in comparison to the TSwind1 and results in a lower optical depth. The shallower optical depths of TSwind2 in comparison with TSwind1 causes its emission to be largely driven by the wind ejecta component, which is uniform along different viewing angles. On the other hand for TSwind1 the dynamical ejecta contribute relatively more to the emission and hence shows a stronger angular dependence.

As was noticed in Sec. \ref{sec:results_morphology} the early blue peak was missing from the TSwind1 model. This peak is retained again here in the TSwind2 case.

\subsection{Estimated light curves}
\begin{figure*}
    \centering
    \includegraphics[width=0.9\columnwidth]{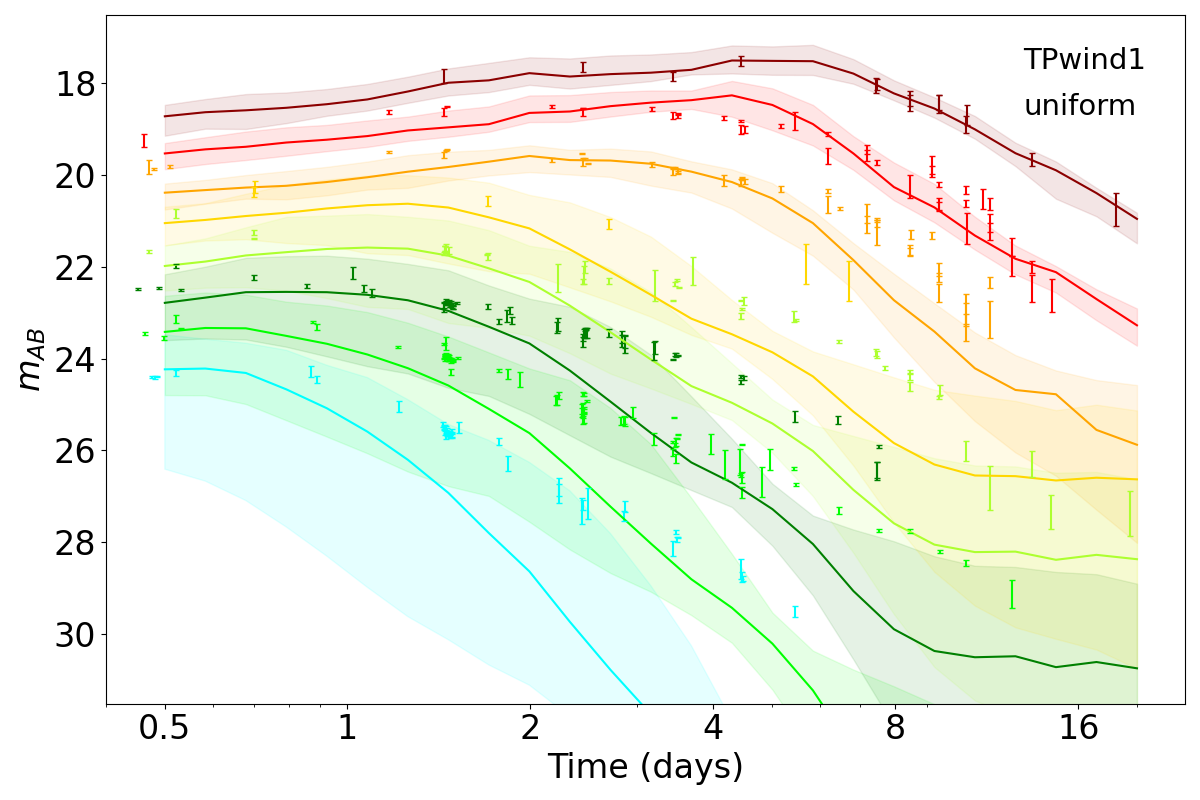}
    \includegraphics[width=0.9\columnwidth]{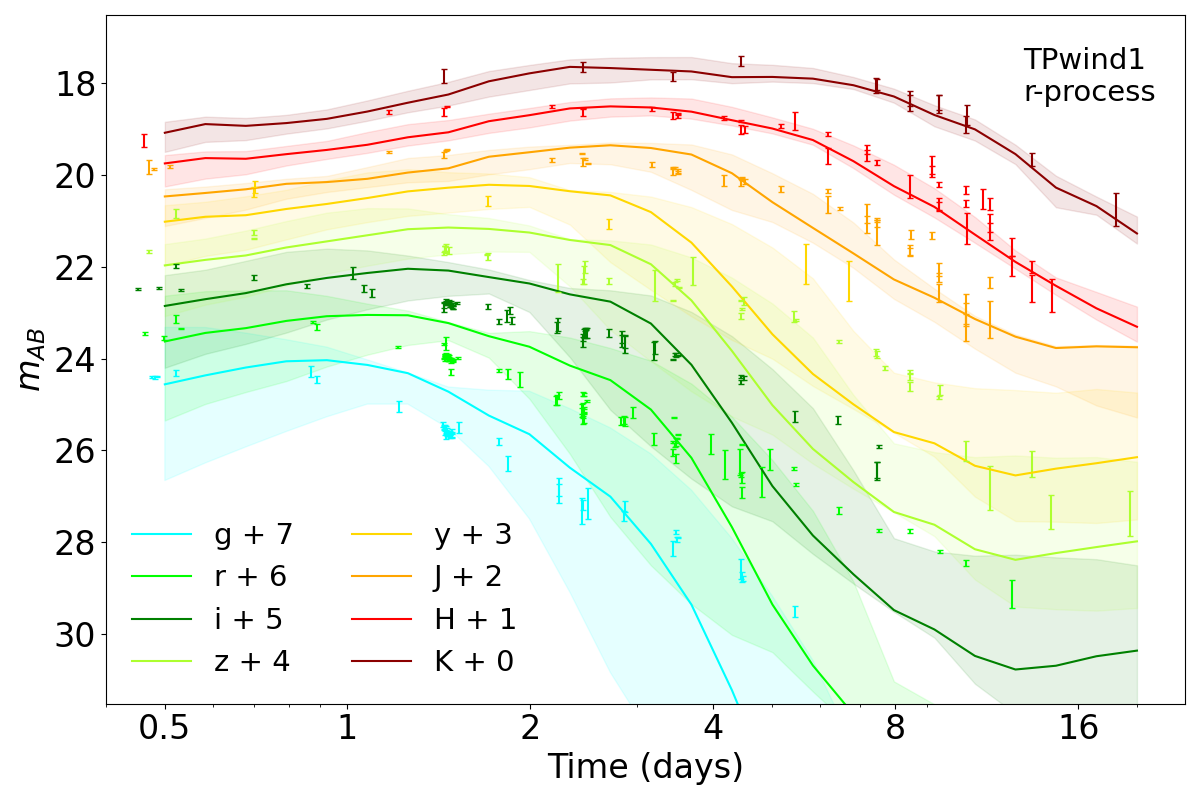} \\
    \includegraphics[width=0.9\columnwidth]{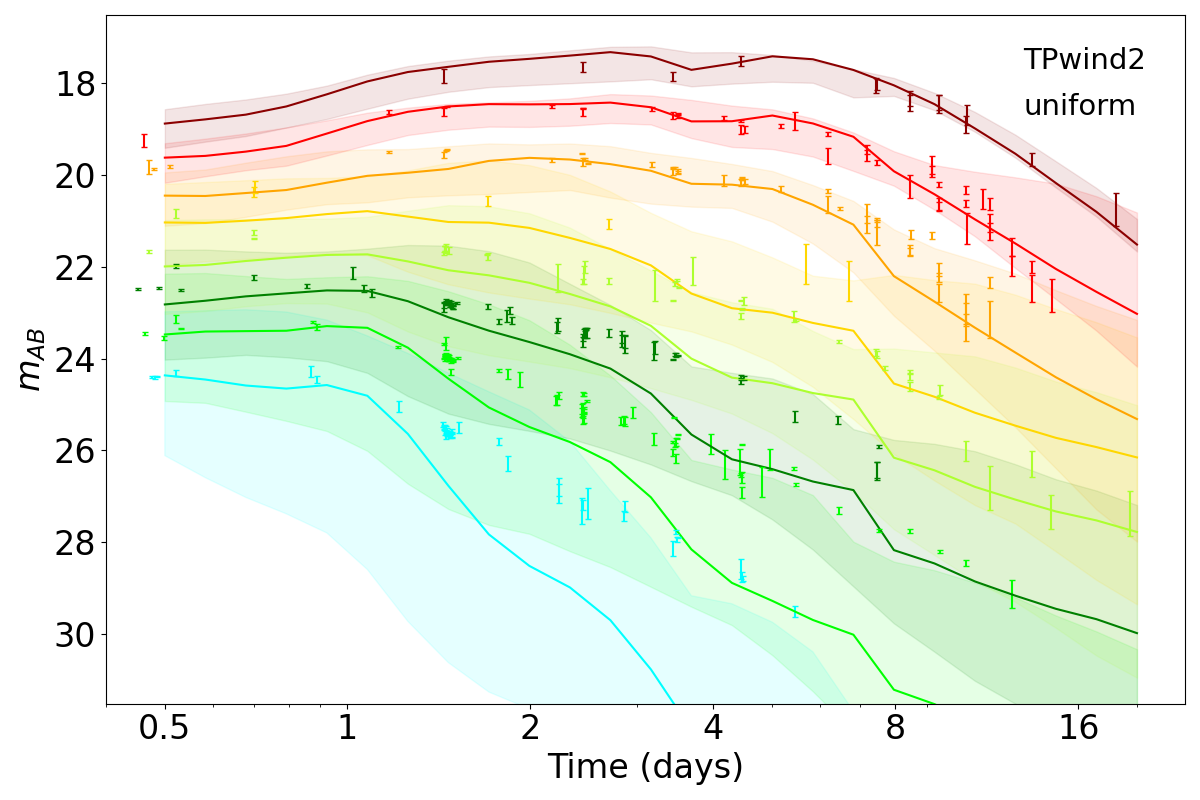}
    \includegraphics[width=0.9\columnwidth]{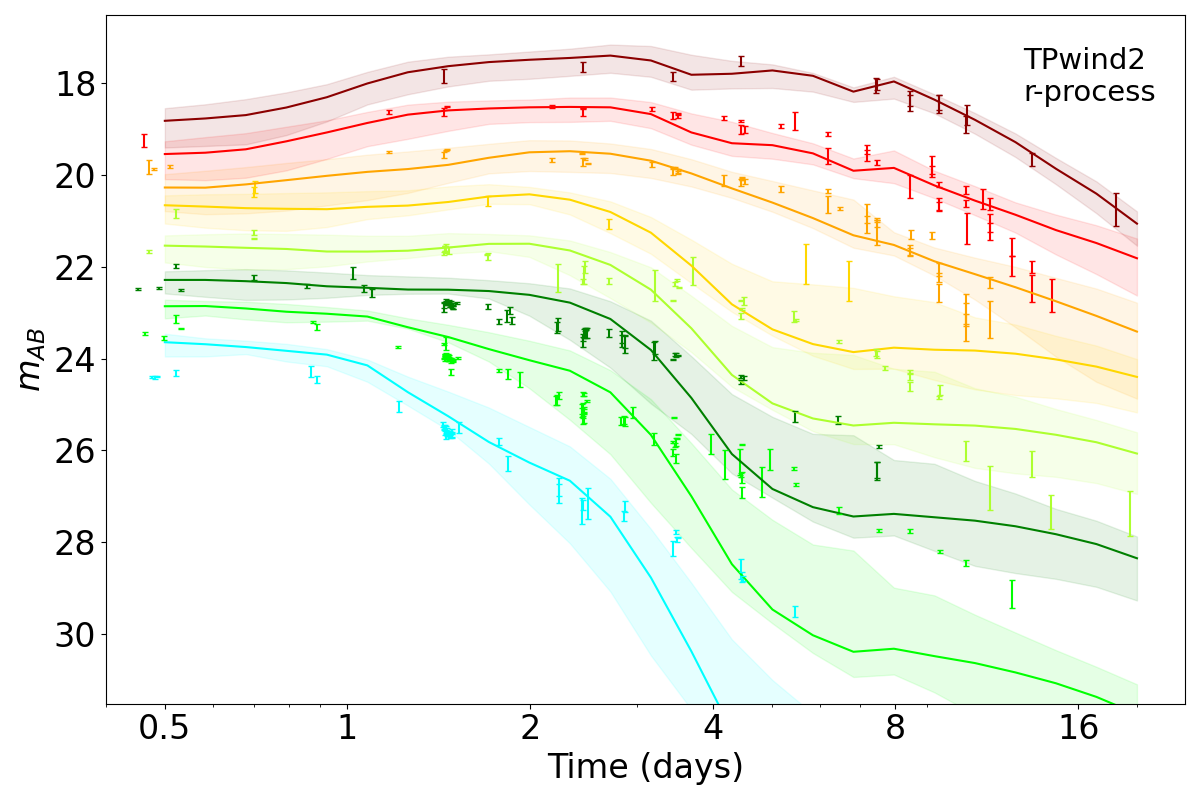} \\
    \includegraphics[width=0.9\columnwidth]{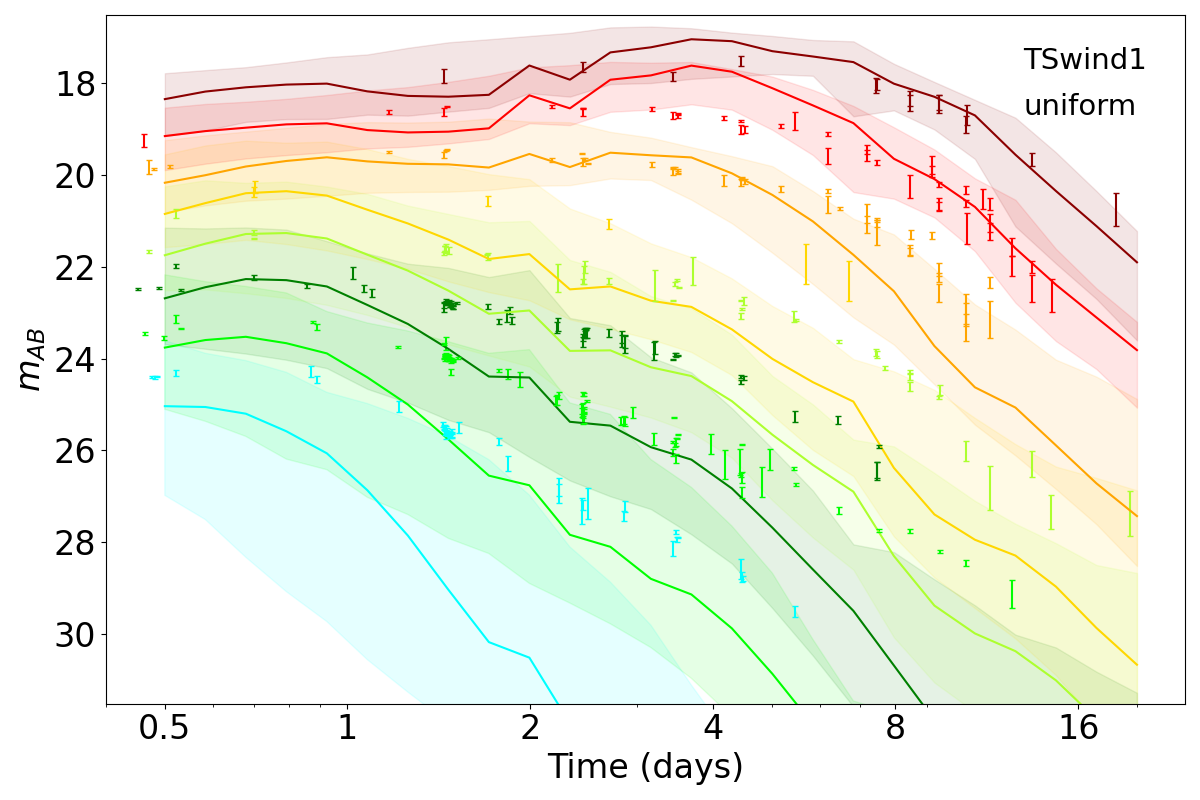}
    \includegraphics[width=0.9\columnwidth]{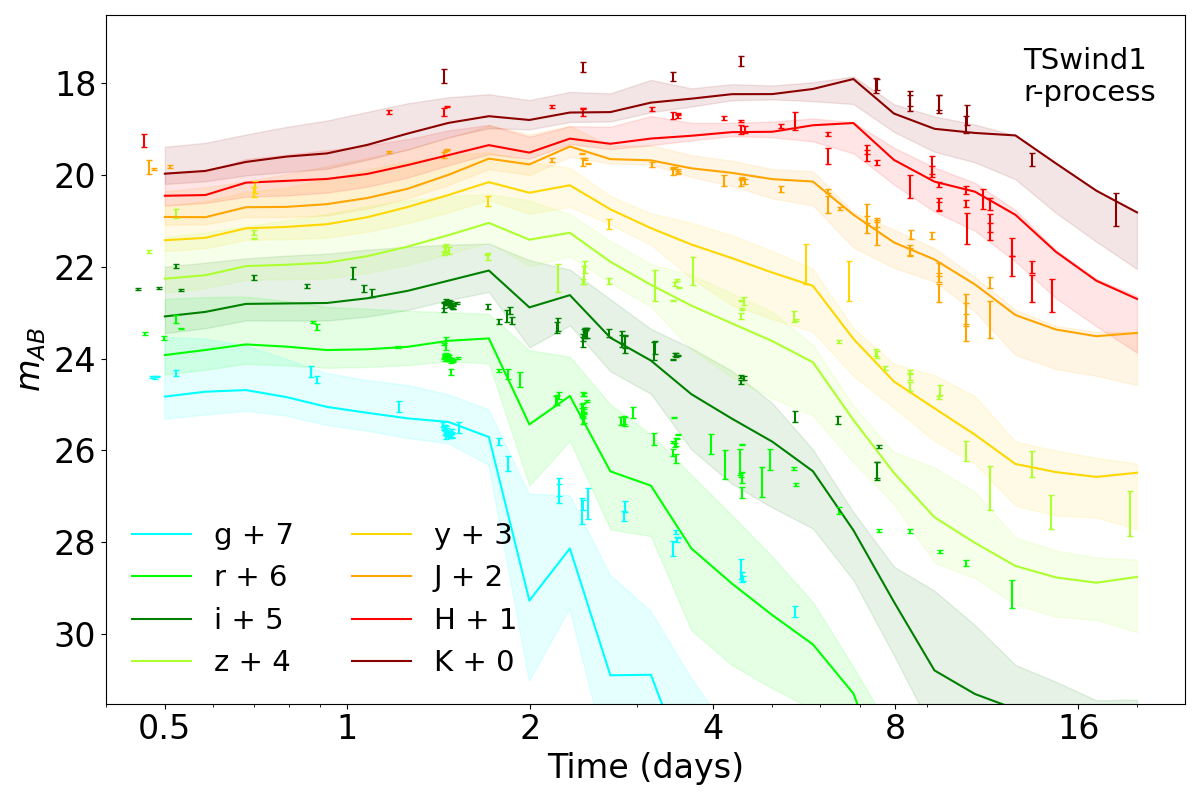} \\
    \includegraphics[width=0.9\columnwidth]{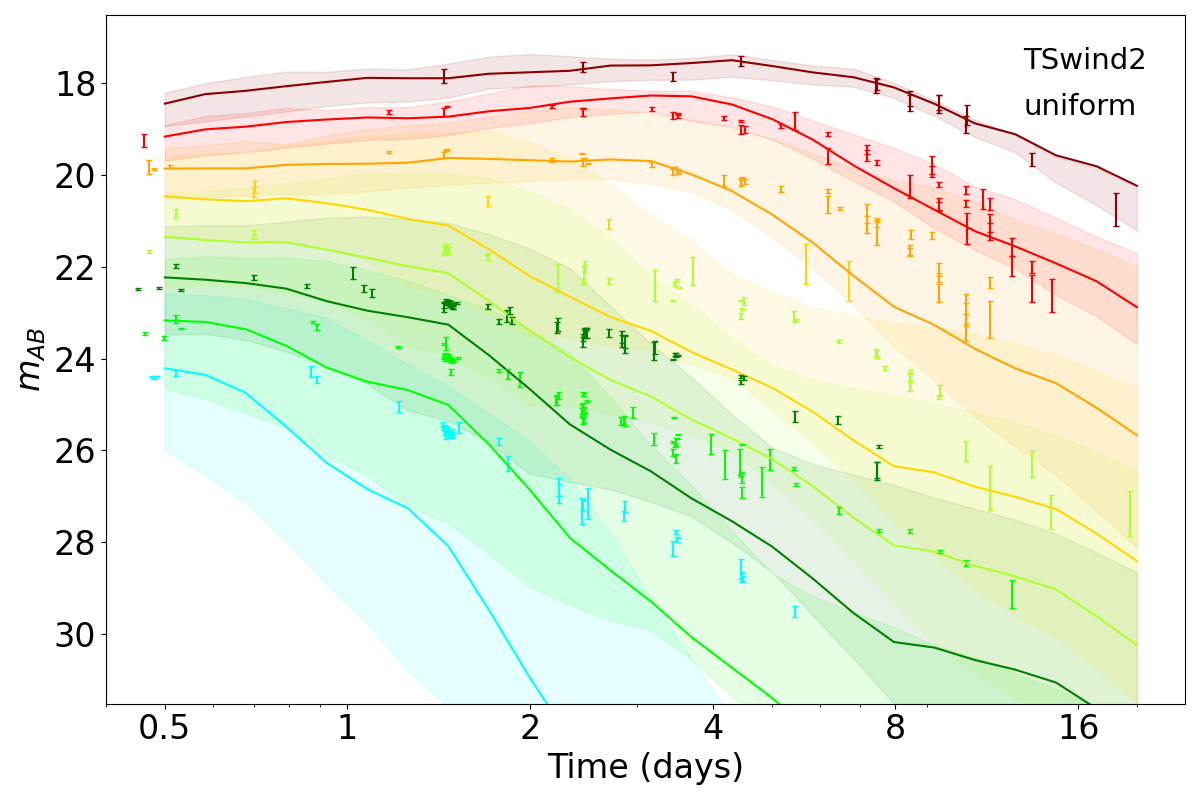}
    \includegraphics[width=0.9\columnwidth]{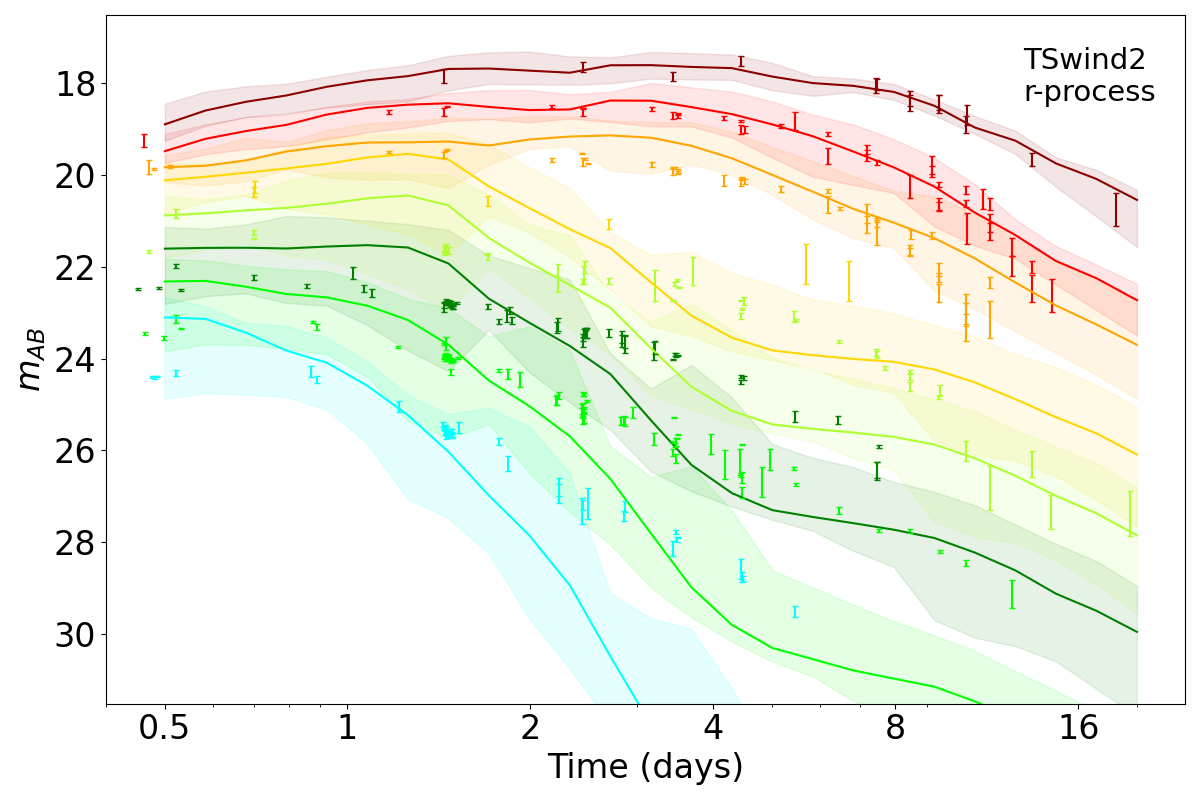}
    \caption{Posterior predicted LCs for AT2017gfo, compared with AT2017gfo observations. Points and error bars denote the observations, with each color denoting a different filter band. Following \cite{Ristic22}, the solid curves and shaded intervals show the median and 90\% credible expected LC, deduced by fitting our models to these observations. The left panel of figures denotes our unconstrained models; the right panels require each outflow to be consistent with Solar \emph{r}-process abundance (i.e., we also adopt a prior on $m_w/m_d$ such that the ejected material; see Sec. \ref{sec:method_ejecta_parameter_inference}). \emph{First row}: TPwind1; \emph{second row}: TPwind2 \cite{Ristic22}; \emph{third row}: TSwind1; \emph{fourth row}: TSwind2.}
    \label{fig:lc}
\end{figure*}

Given the substantial flexibility that we allow in our kilonova model families, each of which has five free parameters (two masses, two velocities and the viewing angle), unsurprisingly, our models can find a good fit for EM observations of GW170817's kilonova LC (taken from \cite{2017ApJ...851L..21V}, which compiles data from \cite{2017Sci...358.1570D, 2017Natur.551...75S, Tanvir_2017, 2017ApJ...848L..17C, 2017Natur.551...64A, 2017Natur.551...71T, 2017ApJ...848L..24V, 2017Natur.551...67P, 2017Sci...358.1559K, 2017Sci...358.1574S, 2017PASJ...69..101U}). Figure \ref{fig:lc} shows the inferred LCs deduced with each of our models. As previously \cite{Ristic22} and unless otherwise noted henceforth, we adopt a strong angular prior based on afterglow observations \cite{2019NatAs...3..940H, Tanvir_2017, 2017Natur.551...71T, 2018Natur.561..355M, Evans_2017, Troja_2020, 2017ApJ...848L..21A}.

In all cases, our unconstrained models fit the observations reasonably well, albeit less well for the bluest bands. Consistent with prior paper \cite{Ristic22}, our posterior predictions for the bluest bands are consistently fainter and more rapidly decaying apart from the TPwind1 case, which fits the $\sim$1 day peak of $g$ band very well. Lacking large residuals, however, these figures provide relatively little insight into the overall goodness of fit. In Table \ref{table:likelihood_ID}, we report the peak likelihood identified overall in each analysis ($\ln L_{\rm max}$, a measure of the goodness of fit of our unconstrained five-parameter kilonova model; and the likelihood associated with the maximum a posteriori (MaP) parameters [$\ln (L\times p)$], a goodness of fit measure. which requires consistency with the prior (either uniform or \emph{r} process). For each morphology and wind composition, the two estimates reported for $\ln L_{\rm max}$ derived from our unconstrained and \emph{r} process constrained analyses are relatively consistent as expected: both measure quality of fit of the unconstrained model. Keeping in mind differences in $\ln L$ of order unity correspond to odds ratio changes of order $e^1$, this table shows that the TSwind1 models fit the data very poorly, but that the remaining unconstrained models all fit the data reasonably well. When the \emph{r}-process prior is imposed, we disfavor the TPwind1 and TPwind2 models slightly in favor of TSwind2.

\begin{table}
    \caption{Peak likelihood identified in parameter inference for GW170817. The mass ratios for the cases where the $r$-process prior was used during parameter inference were peaked at $m_w/m_d = 13.90$ and $1.76$ for the wind1 and wind2 models, respectively. The fit likelihoods correspond to the LCs shown in Fig. \ref{fig:lc}.}
    \begin{tabular}{@{\extracolsep{15pt}}l c c c c @{}}
        \hline \hline
        \multirow{2}{*}{Model} & \multirow{2}{*}{Prior} & \multirow{2}{*}{$\ln L_{\rm max}$} & \multicolumn{2}{c}{Fit likelihood (MaP)} \\
        \cline{4-5} & & & $\ln L$ & $\ln (L\times p)$ \\ \hline
        TPwind1 & uniform           &  2.41 &  --0.15 & 6.63 \\
        TPwind1 & \emph{r} process  &  1.62 &  --0.97 & --2.20 \\\hline
        TPwind2 & uniform           &  0.61 &  --0.62 & 6.77 \\
        TPwind2 & \emph{r} process  &  0.49 &  --1.18 & 4.23  \\ \hline
        TSwind1 & uniform           &    --9 & --11.22 & --4.05 \\
        TSwind1 & \emph{r} process  & --8.78 & --14.32 & --11.26 \\ \hline
        TSwind2 & uniform           &  2.84 &   2.38 & 9.09 \\
        TSwind2 & \emph{r} process  &  2.81 &   1.93 & 6.77 \\
        \hline \hline
    \end{tabular}
    \label{table:likelihood_ID}
\end{table}

\subsection{Discussion of inferred parameters}
\begin{figure*}
    \centering
    \includegraphics[width=\columnwidth]{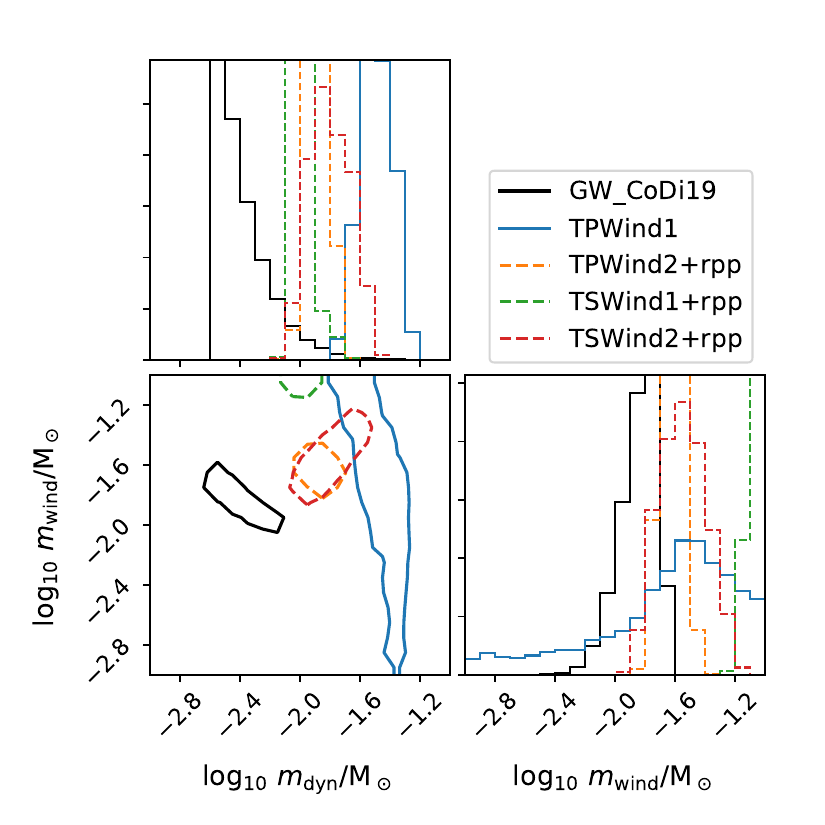}
    \includegraphics[width=\columnwidth]{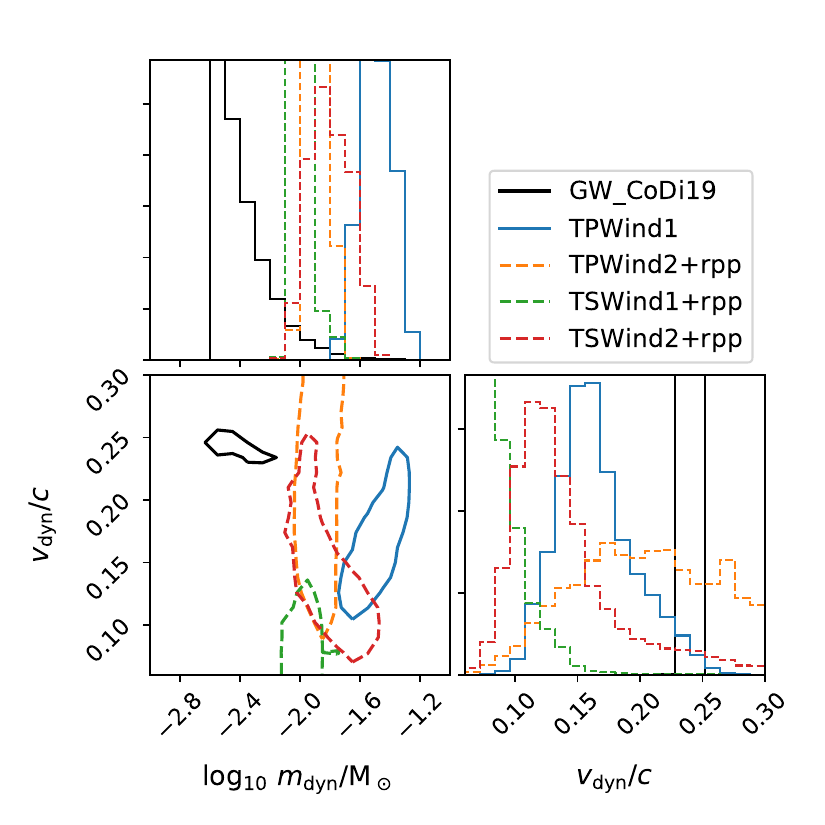}\\
    \caption{
    \textbf{Morphology and composition dependent parameter inference}. This figure shows parameter inference for ejecta properties ($m_d/{\rm M}_\odot$, $v_d/c$, $m_w/{\rm M}_\odot$, $v_w/c$) derived from GW170817 and AT2017gfo. Colored solid and dotted curves show purely electromagnetic inferences derived using different families of two-component LC surrogate models and a prior on source inclination. Different solid colors indicate different choices of outflow and composition with a uniform logarithmic prior on ejecta masses, while colored dotted curves show inferences with the corresponding morphology/composition pairs while also requiring consistency with the Solar \emph{r}-process abundance pattern. The black solid and dotted curves by contrast show three estimates inferred solely from GW measurements, combined with forward models for the outflow. The legend indicates the morphology pair (e.g., TP = Torus and Peanut), while the integer indicates the wind composition (1 refers to the wind1 sequence and so on).  The inferred wind (and thus total) mass depends substantially on the assumed wind composition and morphology. The two-dimensional contours correspond to 90\% credible intervals.
    }
    \label{fig:modified_pe_comp}
\end{figure*}

Despite similarity in inferred LCs, performing ejecta parameter inference with LC surrogate models from alternate morphologies and compositions unsurprisingly produces different inferences for ejecta parameters.
In Fig.~\ref{fig:modified_pe_comp} we illustrate parameter inference for AT2017gfo based on surrogate models for different ejecta models with and without including \emph{r}-process priors. The inferred ejecta properties depend notably on the choice of the outflow's morphology and composition. For example, for our spherical models, the wind ejecta mass is much smaller than the dynamical ejecta mass, and the (dominant) dynamical ejecta mass is strongly constrained.

However, despite broadening our model space, our ejecta inferences remain inconsistent with conventional prior expectations about ejecta from binary mergers, suggesting that persistent modeling systematics remain.

\subsection{Discussion}
\label{sec:results_discussion}

We find that both the alternative morphologies and compositions have a significant impact on emission at $t=1 ~\unit{day}$ and later in the $g$-, $y$-, and $K$-bands, and for emission in the equatorial direction. Reassessing GW170817, we find modestly different conclusions about ejecta parameters, conditioned on outflow morphology and composition. The alternate morphologies help alleviate the tension between observational and theoretical ejecta properties. However, the systematics remain insufficient to completely reconcile our interpretation of the ejecta with our prior expectation about the material ejected from merger based on GW-only inference and theoretical expectations for merger ejecta.

After exploring a range of outflow morphologies, compositions, ejecta masses, and ejecta velocities, we find that the inferences developed from our two-component ejecta models continue to exhibit tension both with prior expectations of ejecta from the merger, using its GW-deduced parameters, and also with the observed LCs themselves missing a blue component.

Our investigations here have omitted two directions, which deserve subsequent attention. First, we have adopted very sharp initial compositions $Y_e$. Full numerical simulations have consistently suggested peaked $Y_e$ distributions with extended tails \cite{ foucart2022general}. A small contribution of suitable material could help explain the bright extended blue component missing in some of our LCs.
Second, and in a similar vein, we have restricted to a two-component model; an additional third small blue component enhancer, previously attributed to magnetically-driven winds could help improve our fit as invoked in previous paper \cite{Nicholl21}. The recently observed kilonova and long gamma-ray burst, GRB211211A, has indicated the presence of an additional thermal component powered by either a GRB jet or a central magnetar \cite{troja2022GRB211211A, yang2022GRB211211A, rastinejad2022GRB211211A, mei2022GRB211211A}. Such a central power source has been proposed such as a long-lived magnetar, a black hole powered by accretion disk or a jet cocoon for previous observations as well \cite{yu2013bright,metzger2014optical,arcavi2018first,matsumoto2018macronova,Piro_2018,li2018powered,2019ApJ...880...22W,ai2021binary}. Our current study excludes such a source to power the emission; however, based on the findings of those studies we anticipate that adding a central engine could help alleviate the early blue-component mismatch and should be investigated further in future work.

Further, we notice that our Gaussian process modeling produces occasional glitches in the surrogate models. These glitches occur only for the wind1 composition models and affect the quality of the predicted LC largely at times beyond 4 days. This error-prone training may get resolved by developing a hyperparameter selection method, which carefully restricts the hyperparameters to preceding parameter neighborhood.

Our inferences of ejecta properties are in modest tension with the expected ejecta properties, assuming GW-derived binary properties and NR-calibrated models for the outflows expected from different binary mergers. However, while we performed an analysis only of GW170817 where these NR-calibrated models are indeed most reliable, at larger mass ratios these models disagree more substantially \cite{henkel2022study}. 
These tensions could be challenging to resolve as a part of joint multimessenger GW-EM PE for generic neutron star mergers.

\section{Conclusions}
\label{sec:conclusions}

In summary, we have developed families of surrogate models for kilonova LCs resulting from binary neutron star merger events. As previously done, using our Gaussian process interpolation applied to an actively-extended simulation archive we develop surrogates for different ejecta configurations: three additional wind morphology and composition combinations in a two-component (dynamical and wind) ejecta approach. Our suite of four kilonova surrogate models now covers multiple outflow configurations comprising toroidal dynamical ejecta, and peanut and spherical shaped wind ejecta with two compositions. These broad set of surrogate models can be employed for parameter inference of ejecta components when kilonovae observations are made in the future and will assist in determining the physics of these events. These kilonovae simulations and their surrogates are available at a Zenodo repository\footref{note1} and the surrogates alone in a GitHub repository \footref{note2}.

As expected, we find that wind morphology and, to a lesser extent, composition (within our constrained set of compositions) have substantial impacts on the outgoing emission's time, frequency, and angular dependence.
To assess the impact of these differences, we calculate the likely ejecta properties associated with GW170817's kilonova, conditioned on the assumption that the ejecta and radiation exactly reproduce one of our assumed ejecta models. We find that the choice of model most strongly impacts the inferred ejecta masses. None of our inferences strongly constrains the ejected wind velocity.

To better understand the context of these calculations, we compare and contrast them with the inputs and outputs expected from GW170817's merger.  On the one hand, we compare our inferred ejecta masses with the expected ejecta masses deduced from GW170817's GW-constrained masses, using several contemporary estimates for the ejecta.  While the EM- and GW-deduced ejecta can be reconciled for some ejecta models, allowing for substantial fit and EOS uncertainty, we generally see considerable tension in the inferred ejecta mass, with the EM-inferred masses generally larger than the GW-deduced ejecta masses when assuming a single plausible equation of state.  Such tension between GW-deduced (and hence numerical-relativity informed) ejecta masses has been repeatedly highlighted in the literature \cite{2019MNRAS.489L..91C, 2020Sci...370.1450D, Pang2022NMMA,2020PhRvD.101j3002K, dietrich2017modeling, 2019MNRAS.489L..91C, henkel2022study} (a recent paper indicates the possibility of relieving this tension with the use of new heating rate fitting formulas \cite{BullaPOSSIS2023,rosswog2022heavy}). Our corroboration of this finding, even using state-of-the-art ejecta calculations, suggests the resolution of this discrepancy must invoke other sources of systematic error, for example in the ejecta initial conditions.
On the other hand, we repeat our ejecta mass inferences while requiring the ejected \emph{r}-process mass abundances to be consistent with Solar \emph{r}-process patterns. While this  constraint sharply reduces the range of ejecta masses allowed, it does not significantly change any of the aforementioned conclusions: the EM-deduced ejecta masses remain large, while EM-constrained ejecta velocities remain weakly constrained.  Our \emph{r}-process constraints on $m_w/m_d$ remain qualitatively consistent with the corresponding ratios expected from forward-modeling ejecta starting from GW inferences about binary masses.

Our kilonova generation work has limitations in the following ways, which should be addressed in future work. The two-ejecta component model is known to account for the lanthanide rich and lanthanide free outflow materials to a considerable degree. However, the LC being substantially different for differing morphologies implies that we would need a more exhaustive calculation of ejecta LCs with other morphologies or account for that with a third component. Hence, adding a third component or a continuous ejecta model could cause a shift in LCs.
Similarly, our present and previous work suggests that adjusting our assumptions about the composition, both in median and distribution, can also notably impact the LCs. We have also highly simplified the underlying nuclear physics, both in our assumptions about heating \cite{2021ApJ...918...44B, 2021ApJ...906...94Z, Barnes_2016} and in using $Y_e$ as a proxy for the detailed isotopic composition. Finally, our results depend on our present understanding of pertinent opacities, which, while dramatically improved over prior work, does not yet include the latest actinide opacities, let alone non-LTE physics at late times. 

For technical convenience, our discussion of the tension between our direct inferences about ejecta from EM and secondhand conclusions about ejecta derived from GW inference of GW170817 adopted a fixed nuclear equation of state (APR4) to predict ejecta properties from binary properties.  By eliding uncertainty in the nuclear equation of state in our figure, we somewhat underestimate the systematic uncertainty in ejecta estimates derived from GW measurements, artificially strengthening the apparent tension between GW- and EM-deduced ejecta properties.

\acknowledgements
The authors thank Erika M. Holmbeck and Matthew R. Mumpower for stimulating discussions throughout the development of this research. A.K. and M.R. acknowledge support from National Science Foundation (NSF) Grant No. AST-1909534. R.O.S. acknowledges support from NSF Grant No. AST-1909534, NSF Grant No. PHY-2012057, and the Simons Foundation. A.B.Y. acknowledges support from NSF Grant No. PHY-2012057. R.T.W., O.K., E.A.C., C.L.F., and C.J.F. were supported by the U.S. Department of Energy through the Los Alamos National Laboratory (LANL). This research also used resources provided by LANL through the institutional computing program. Los Alamos National Laboratory is operated by Triad National Security, LLC, for the National Nuclear Security Administration of U.S. Department of Energy (Contract No. 89233218CNA000001). Research presented in this paper was supported by the Laboratory Directed Research and Development program of Los Alamos National Laboratory under Project No. 20190021DR.

\bibliography{bibliography,LIGO-publications,gw-astronomy-mergers-ns-gw170817}

\end{document}